\documentclass{article}

\usepackage{wrapfig}
\usepackage{multicol}
\usepackage{graphicx,caption,subcaption,color}
\usepackage{amsmath}
\usepackage{multirow,multicol}

\usepackage{arxiv}

\usepackage[utf8]{inputenc} % allow utf-8 input
\usepackage[T1]{fontenc}    % use 8-bit T1 fonts
\usepackage{hyperref}       % hyperlinks
\usepackage{url}            % simple URL typesetting
\usepackage{booktabs}       % professional-quality tables
\usepackage{amsfonts}       % blackboard math symbols
\usepackage{nicefrac}       % compact symbols for 1/2, etc.
\usepackage{microtype}      % microtypography
\usepackage{lipsum}		% Can be removed after putting your text content
\usepackage{graphicx}
\usepackage{natbib}
\usepackage{doi}

\title{The radio emission of the core in Fanaroff-Riley type II radio galaxies}

\date{} 					% Or removing it

\author{F.~Mazoochi$^1$,  H.~Miraghaei$^2$\thanks{\href{Corresponding author Email: h.miraghaei@maragheh.ac.ir}{Corresponding author Email: h.miraghaei@maragheh.ac.ir}},  N.~Riazi$^1$\\
$^1$Department of Physics, Shahid Beheshti University, Tehran, Iran\\%
$^2$Research Institute for Astronomy and Astrophysics of Maragha (RIAAM), University of Maragheh, Maragheh, Iran}

\renewcommand{\headeright}{\textit{F. Mazoochi \& H. Miraghaei \&  N. Riazi}}
\renewcommand{\undertitle}{}
\renewcommand{\shorttitle}{}

\hypersetup{
pdftitle={A template for the arxiv style},
pdfsubject={q-bio.NC, q-bio.QM},
pdfauthor={David S.~Hippocampus, Elias D.~Striatum},
pdfkeywords={First keyword, Second keyword, More},
}
\begin{document}
\twocolumn[{
\maketitle

\begin{abstract}
We study the radio power of the core and its relation to the optical properties of the host galaxy in samples of high excitation (HERG) and low excitation (LERG) Fanaroff-Riley type II (FRII) radio galaxies. The radio galaxy sample is divided into two groups of core/non-core FRII, based on the existence of strong, weak or lack of single radio core component. We show that FRII LERGs  with radio emission of the core have significantly higher [O III] line luminosities compared to the non-core LERG FRIIs. There is no significant difference between the hosts of the core and non-core FRIIs of LERG type in galaxy sizes, concentration indices, star formation rates, 4000-\AA\ break strengths, colours, black hole masses and black hole to stellar masses. We show that the results are not biased by the stellar masses, redshifts and angular sizes of the radio galaxies. We argue that the detection of higher [O III] luminosities in the core FRIIs may indicate the presence of higher amounts of gas, very close to the AGN nucleus in the core FRIIs compared to the non-core FRIIs or may result from the interaction of the radio jets with this gas. The core and non-core FRIIs of the HERG type show no significant differences perhaps due to our small sample size. The effect of relativistic beaming on the radio luminosities and the contribution of restating AGN activity have also been considered.
\end{abstract}

% keywords can be removed
}]

\section{Introduction}
The presence of supermassive accreting black holes at the centers of massive galaxies is associated with significant radiation at all wavelengths across the electromagnetic spectrum \citep{netzer2013physics}. This is a main feature that distinguishes galaxies with active nuclei from normal, mostly star-forming ones. In terms of radio emission, active galactic nuclei (AGN) are commonly divided into two classes of radio-quiet and radio-loud AGN. Radio-loud AGN defined by their high radio emissions, exhibit either a compact or large scale relativistic jet and lobes \citep{wilson1994difference}. In contrast, radio-quiet AGN show no or very weak relativistic jets or winds \citep{kharb2020probing}. 

The morphologies of very extended jets and lobes in radio-loud AGN are classified into two main different classes known as Fanaroff-Riley (FR) dichotomy \citep{fanaroff1974morphology}. In the first class of FR radio galaxies (FRI), the distance between the peak of the emission
on the opposite sides of the radio source was smaller than half of the total size of the radio source. In the second
class of FR radio galaxies (FRII), the peak of the emission locates at the second half of the distance between the
core and the edge of the emission. Based on the definition, FRIIs (edge-brightened) contain weaker radio emissions in their cores and on the contrary, FRIs (edge-darkened) consist of more powerful radio emissions at their centers. Therefore, a prominent characteristic of FRII jets is the lower ratio of core to total radio luminosity compared to FRI radio galaxies. Recently, compact radio sources has
been added to this classification as a third class of radio galaxies as FR0 by \citet{baldi2015pilot}. The FR0s consist of the maximum amount of core to total radio luminosity ratio. They are core dominated radio galaxies at a higher level compared to FRIs and are distinguished from FRIs by the lack of substantial extended emission. Therefore, there is, roughly, a one-to-one map between the morphological class of radio galaxies and the level at which the core dominates the total radio emission of a galaxy.
The origin of compact radio sources such as FR0s and their relations to the extended sources are still under debate in the literature. Investigation of the proper motion and spectral analysis of compact steep spectrum (CSS) and peaked spectrum (PS) radio sources suggest that these sources are young objects probably at the beginning of their jet explosions. They could evolve into larger radio sources or they may be unable to launch large-scale jets due to their dense environments or their intrinsic properties such as black hole mass and spin (see \citet{o2021compact} for a review).

In addition to the main classification, radio galaxies with more complex structures have been detected.
There are Hybrid radio sources \citep{wiita2000extragalactic} which are FRI on one side and FRII on the other side, Narrow- and Wide-Angle-Tailed \citep{owen1976radio}, Head-Tailed \citep{owen1976radio} and Double-Double sources \citep{schoenmakers2000radio} which are considered as subclasses to the main FR classification.
By improvement in the sensitivity and resolution of current and future radio surveys such as Low Frequency Array (LOFAR) and Square Kilometre Array (SKA), 
more details about the morphology of radio galaxies are discovered and sources with probably even more complex structures    
are detected \citep{mingo2019revisiting}.

In order to find the origin of morphological differences observed in radio galaxies, the relation between the morphology and AGN host galaxy properties or environments has been investigated. Numerous studies have shown that environmental factors are able to affect the jet morphology \citep{hill1991change,lin2010populations, miraghaei2017nuclear, massaro2019deciphering}. The FRIs are frequently in denser environment than the FRIIs. A possible explanation for jet disruption is the interaction of the jet with environmental matter. Therefore, FRIs are more likely to get disrupted sooner than the FRIIs. In terms of host galaxy properties, the main results indicate that FRIs are typically hosted by elliptical galaxies with higher mass surface density which is consistent with the jet disruption scenario driving the FR dichotomy \citep{bicknell1994relativistic,laing2002dynamical,miraghaei2017nuclear}.

In this study, we explore core radio emissions in a sample of radio-loud AGN. Our motivation is to find out the characteristics of galaxies which lead to the production of small scale radio jets. In addition, investigation of the radio emission of the core can provide an important insight into the radio-loud/radio-quiet dichotomy. For this aim, we use a sample of FRII radio galaxies. 
Unlike FRIs and as discussed above, FRIIs are not considered as core dominated radio galaxies. The radio emissions of these
galaxies peak at the edge of the radio lobes where a quite secure distance separates the radio emission of the
core and the two-sided lobes. Therefore, they provide the most efficient core/lobe classification for investigation  
of radio emissions of AGN at kpc scale. Our preference for using FRII radio galaxies instead of FR0s in which 
the radio emission only have a core component, is because i) the existence of extended emission in 
FRIIs indicates that the jets expand perpendicular or make a wide angle to the line of sight. Therefore, the contribution of the core emission from the extended component which is due to the projection effect is minimal. In addition, the analysis is less contaminated by the relativistic beaming from the jet due to the viewing angle however reorientation of the radio jets has been reported in the literature \citep{kharb2006radio,kharb2014very,hernandez2017restarting}. This may impose error to this analysis; ii) the unknown nature of
FR0s in terms of e.g. their black hole spins or other characteristics which may influence 
the radio emission of the core, does not influence our results since FRIIs can display either the extended 
or core radio emission; iii) As previous studies showed that the environment may play role in FR 
morphological dichotomy, limiting the radio sample in this analysis to FRIIs, removes or reduces the potential biases
caused by the environment. In addition, it even eliminates other factors such as magnetic field and kink instability that can control the jet morphology \citep{tchekhovskoy2016three}.

In general, the radio emission of the core at kpc scale can originate from either the star formation or AGN activity of galaxies. The radio emission from star formation comes from the thermal free-free emission of H II regions \citep{terzian1965radio} or non-thermal synchrotron emission of cosmic rays associated with supernovae which accelerate into the galactic magnetic field \citep{condon1992radio}. The latter produces a tight Far-Infrared radio correlation in the continuum spectrum of star forming galaxies \citep{condon1991correlations,gurkan2018lofar}. The radio emissions from AGN activity can be from compact or low-powered radio jets \citep{falcke2000radio,kharb2019curved,jarvis2019prevalence} or winds \citep{cecil2001jet,irwin2003giant,hota2006radio}
or may represent large scale jets at the beginning of explosions. 

In this study, we compare the properties of host galaxies of FRIIs which have core radio emissions with those with no or weak radio emission of the core to find out the similarities and differences between them. We benefit from using samples of the same morphological classes to get the most robust results on the origin of the radio emission of the core as well as some or any clues to the radio-loud/radio-quiet problem. Throughout this paper, we assume a $\Lambda CDM$ cosmological model with the following parameters: $\Omega_m=0.3$, $\Omega_{\Lambda}=0.7$ and $H_0=100hkm^{-1}Mpc^{-1}$ where $h=0.70$.

\section{Radio sample and optical characteristics}\label{sec2}
\label{sec:pops}
\subsection{Radio data}
\label{sec2.1}
%\textbf{2.1\ Radio data}
The FRII radio galaxy sample has been extracted from \citet{miraghaei2017nuclear}. The sample provides 646 FRII sources with secure identifications of their host galaxies in the radshift range  $0.03<z<0.6$. The catalogue has been originally constructed by the mophological classification of the extended radio source sample reported in \citet{best2012fundamental}. \citet{best2012fundamental} have cross-matched the seventh data release (DR7; \citet{abazajian2009seventh}) of the Sloan Digital Sky Survey (SDSS; \citet{york2000sloan}) with the Faint Images of the Radio Sky at Twenty centimeters (FIRST; \citet{becker1995first}) and the National Radio Astronomy Observatory (NRAO) Very Large Array (VLA) Sky Survey (NVSS; \citet{condon1998nrao}) to produce a large sample of galaxies with their radio source counterparts. They listed the total and core radio luminosity for each galaxy at 1.4 GHz. The identification of star-forming galaxies and radio AGN has also been performed. The radio AGN sample has then been divided into low excitation radio galaxies (LERG) or high excitation radio galaxies (HERG) according to the optical spectra. 
We have selected FRII radio galaxies in the redshift range $0.03<z<0.3$. This includes 409 FRII radio galaxies. The core radio luminosities as reported in \citet{best2012fundamental} have been extracted from the FIRST data. A higher resolution of the FIRST survey (${\theta}{\sim}$5$''$) in comparison to the NVSS (${\theta}{\sim}$45$''$) helps resolving multi-component radio emissions of AGN and provides a higher precision for the detection of the core component. The component refers to a single detection by the source finding algorithm rather than a single physical radio source which may be described by multiple components. The detection and identification techniques are described by \citet{best2005sample}. Based on this technique, \citet{best2012fundamental} crossmatched the SDSS galaxies with the NVSS catalogue within 3 arcmins. Candidates in NVSS with one match (single-component), two matches (doubles), three matches (triples), four and more are investigated to find out true extended radio galaxies rather than physically unrelated sources. Because of the lower resolution of NVSS compared to FIRST, single-component NVSS sources may display as multi-component sources by the FIRST. To find out these extended radio galaxies, the NVSS single-component sample is crossmatched with the FIRST catalogue. The resultant sample is again reconsidered based on the number of matches (components) found from the FIRST catalogue. In addition, to increase the reliability of NVSS multi-component radio sources as true extended sources, their FIRST counterparts have also been considered. Finally, the physically related radio components are found based on the location of the host galaxy, the flux weighted mean position of the radio components, the angle made by the radio components and the host galaxy or visuall inspection for complicated cases. For all of the objects which are accepted as extended sources, the central FIRST component within 3 arcseconds of the optical SDSS galaxy is extracted as radio core emission. Sources with flux densities above $5 \sigma$ (the local noise level) are considered as true detections. In total, 254 FRIIs show the central FIRST components. The integrated flux over the NVSS component, rather than the FIRST component, is used to extract the total radio emission. This corrects the missing flux density of the FIRST data, resulting from the absence of short spacings. 
\begin{figure}[h!]
	\begin{center}
		\includegraphics[width=\columnwidth]{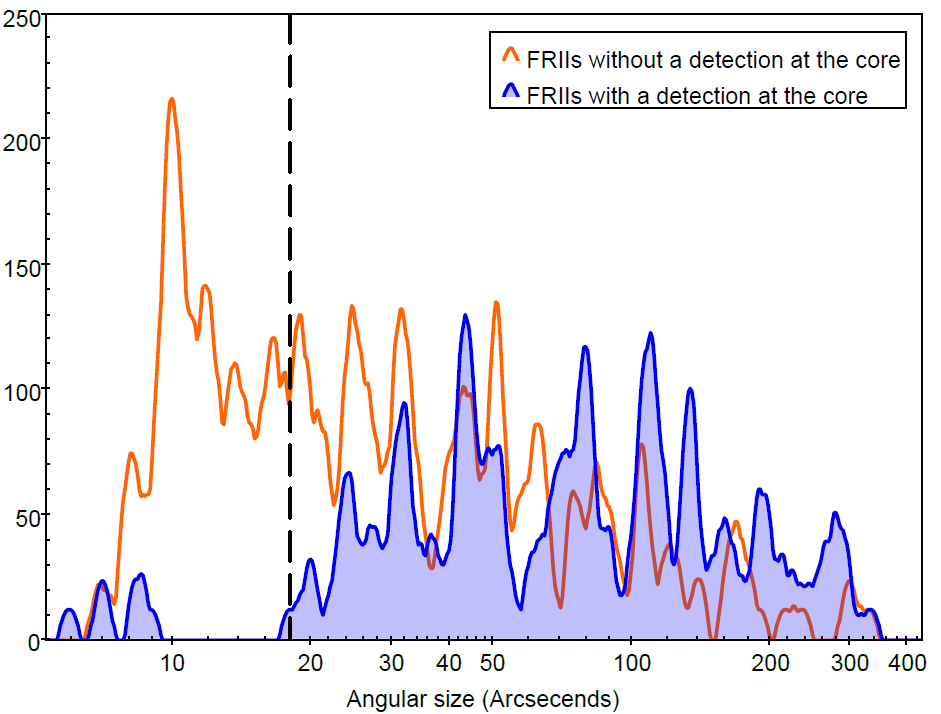}
		\caption{The distribution of angular sizes for the FRIIs with (filled histogram) and without (empty histogram) a detection of a radio core. A cut at 18 arcseconds is applied to both samples to reduce the uncertainties caused by the low resolution of the data. The plot includes 409 FRIIs reported by \citet{miraghaei2017nuclear}. } 
		\label{fig1}
	\end{center}
\end{figure}
The angular sizes of FRII radio galaxies as presented by \citet{miraghaei2017nuclear} show the total sizes of multi-component radio structures including the cores, jets and lobes. Fig. \ref{fig1} shows the distribution of angular sizes of 409 FRII radio galaxies used in this study. FRIIs with a core detection and those without a core detection are presented in blue (filled) and red (empty) colours respectively. A significant fraction of radio galaxies with no detection at the core has small angular sizes while galaxies with core radio detection lie above $\sim$18 arcsec. This definitely shows that sources with small angular sizes are not resolved enough for secure detection of core radio components. Therefore, there may be some sources with core radio emission which are identified as galaxies without radio emission at the core. In order to prevent any bias from this, we need to apply a cut at $\sim$18 arcsecends for the FRII sample. This equals about three times the FIRST resolution (as the beam size of the survey varies between $ 5.4^{\prime \prime} \times 5.4^{\prime \prime}$ and $5.4^{\prime \prime} \times 6.8^{\prime \prime}$ depending on the declination of the sources) which provides a secure identification of the core. 303 FRII radio galaxies remain to use in the rest of the analysis.

We also added the data obtained from the Very Large Array Sky Survey (VLASS) epoch 1 Quick Look imaging at 3 GHz 
\citep{gordon2020catalog}
to calculate the 1.4-3 GHz spectral index of the core component. The radio component above $5 \sigma$ level of the local rms and within 3 arcseconds of the host galaxy are selected as a true detection for the core. The completeness limit of the VLASS is taken to be 3 mJy.

\subsection{Host galaxy properties of the sample}\label{sec2.2}
%\textbf{2.2\ Host galaxy properties of the sample}

The optical properties of galaxies were taken from the value-added spectroscopic catalogues produced by the group
from the Max Planck Institute for Astrophysics and Johns Hopkins University (MPA-JHU, \cite{brinchmann2004physical}). These properties are widely described in \citet{kauffmann2003stellar,kauffmann2003host,kauffmann2003dependence}. For the sake of clarity, we define them in what follows:

\textbf{Stellar mass} ($M_{\star}$) is based on the photometry data, derived from the mass to light ratio \citep{kauffmann2003stellar}.

\textbf{4000-\AA\ Break} ($D_{4000}$) is an approximation for the stellar population age which was introduced by \citet{bruzual1983spectral}.  The 4000-\AA\ break is created by absorption lines displayed around 4000\ \AA \ in the spectrum of the galaxies and is influenced by temperature and metallicity.
In general, the temperature of stellar atmospheres decreases with age while the metal opacity strength increases by age \citep{kewley2006host}. The parameter is small for galaxies with younger stellar populations and large for galaxies with older ones.
We have used the amplitude of the 4000-\AA\ break of the narrow version of the index defined in Balogh et al. (1999):
The ratio of the flux in the red continuum to that in the blue continuum ($D_{4000}=\int_{4000}^{4100}f_{\lambda}d_{\lambda}/\int_{3950}^{3850}f_{\lambda}d_{\lambda}$).The narrow band width of 100 \AA is assumed.
The parameter is also corrected for the observed contributions of the emission lines in this bandpass \citep{kauffmann2003host}.

\textbf{Radius} ($R_{50}$) which is calculated from Petrosian radius. The Petrosian radius is defined as the radius at which Petrosian ratio equals some particular value. We use $R_{50}$ that contains 50 percent of Petrosian flux in the $r$-band \citep{blanton2001luminosity}.

\textbf{Concentration} ($R_{90}/R_{50}$) is calculated as the ratio of $R_{90}$ to $R_{50}$. $R_{90}$ and $R_{50}$ are radii that contain 90 and 50 percents of Petrosian flux. High concentration index objects have a strong central concentration of lights and faint wings but low concentration index objects nearly exhibit a uniform light distributions \citep{blanton2001luminosity}. This index for early-type galaxies is more than 2.6 but in the other hand for late-type and disk-dominated galaxies is mainly lower than 2.6 \citep{kauffmann2003dependence}.

\textbf{Star formation rate} ($SFR$) is as described in \citet{brinchmann2004physical}. This is expressed in $M_{\odot}/ yr$. 

\textbf{Specific star formation rate} ($sSFR$) which is defined as star formation rate per unit galaxy stellar mass \citep{bauer2007specific}.

\textbf{Colour} ($g-r$) which is the magnitude difference at rest-frame. The value increases as galaxies evolve from blue to red.

\textbf{Half-light surface mass density} ($\mu_{50}$) is computed through this relation: $\mu_{50}=0.5M_{\star}(\pi R_{50}^2)$  where $M_{\star}$ is the stellar mass of the galaxy.

\textbf{Black hole mass} ($M_{BH}$) is obtained from renowned $M-\sigma$ relation that shows a strong correlation between mass of central black hole ($M_{BH}$) and the stellar velocity dispersion ($\sigma_{\star}$). This relation is described as:

 $\log{M_{BH}/M_{\odot}}=8.13+4.02\log{\sigma_{\star}/ 200kms^{-1}}$
 \citep{tremaine2002slope}.

\textbf{O III luminosity} ($L_{[OIII]}/L_{\odot}$) is calculated from the [O III] 5007 emission line. The luminosity is detected with $S/N>2.5$ for a large number of objects.
The emission line luminosities below this signal to noise ratio are replaced by zero value as L$_{O III,min}$ or by 2.5 times of the noise as L$_{O III,max}$.   

\subsection{AGN sample and classification}
\label{sec2.3}
%\textbf{2.3\ AGN sample and classification}
According to the orientation-based unification schemes for AGN, the observational properties of AGN are function of the viewing angle \citep{antonucci1993unified,urry1995unified}. Based on that, the optical continuum in type I AGN are dominated by the non-thermal emission at the galactic nucleus than the host galaxy and the stellar population. Therefore, for investigation of the host galaxy properties, type I AGN are excluded from the main galaxy sample by the automated SDSS classification pipeline.

In terms of type II AGN, Schmitt, Storchi-Bergmann and Cid Fernandes (1999) have carried out a detailed spectral synthesis analysis of spectra of the nuclear regions of 20 nearby type 2 Seyfert galaxies and find out the continuum contribution of the AGN is extremely small ($<$5 percent). \citet{kauffmann2003host} discussed this for a sample of type II AGN selected from the SDSS main galaxy sample and showed that AGN light does not contribute significantly to the continuum emission in type II AGN. They particularly discussed the stellar age, defined by 4000-\AA\ break, and other host galaxy properties defined by the basic galaxy structural properties derived from the SDSS imaging data. Those properties are widely used in this study.

The radio AGN are basically divided into two classes of HERG and LERG according to the relative intensity of high- and low-excitation lines in their optical spectra \citep{hine1979optical,bridle1994deep}. In this study, we used the HERG/LERG classification presented in \citet{best2012fundamental}.
They considered a) the ratios of four high-excitation lines ([O III], [N II], [S II] and [O I]) to the $H_{\alpha}$ and $H_{\beta}$ emission lines, b) the equivalent width of the [O III] emission line and c) the line-ratio diagnostic diagrams from \citet{kewley2006host} and \citet{fernandes2010alternative} to identify high and low excitation radio galaxies. By using the excitation class of AGN, we remove the potential biases caused by this on the results as HERGs and LERGs represent different accretion rates into supermassive black hole (SMBH), and are hosted by different galaxies. HERGs are believed to be in high accretion rate mode, having a geometrically-thin, optically thick accretion disk, strong emission lines and obscuring structure of dusty molecular gas while LERGs are accreted in low accretion rate mode by a geometrically-thick advection dominated accretion flow, having weak narrow optical emission lines and have less or no contamination by obscuring structure (see \cite{heckman2014coevolution} for a review). In total, 38 and 357 FRII radio galaxies are classified as HERGs and LERGs respectively.

\section{Selection of galaxies based on the core emission}
\label{sec3} 

In this section, we classify the FRII sample, based on the core radio luminosities, into two samples of i) FRIIs with medium/strong radio core emission (hereafter core FRIIs) and ii) FRIIs with weak or undetectable radio core emission (hereafter non-core FRIIs), as defined below/in Section \ref{sec3}. Then, we compare the host galaxy properties of these two samples in the following sections. Galaxies with no radio detection at their core are described in Section \ref{sec2.1}. The definition for the galaxies with the very weak radio core is based on the redshift of study and will describe later in this section.

\begin{figure}[h!]
	\centering
	\includegraphics[width=\columnwidth,height=6cm]{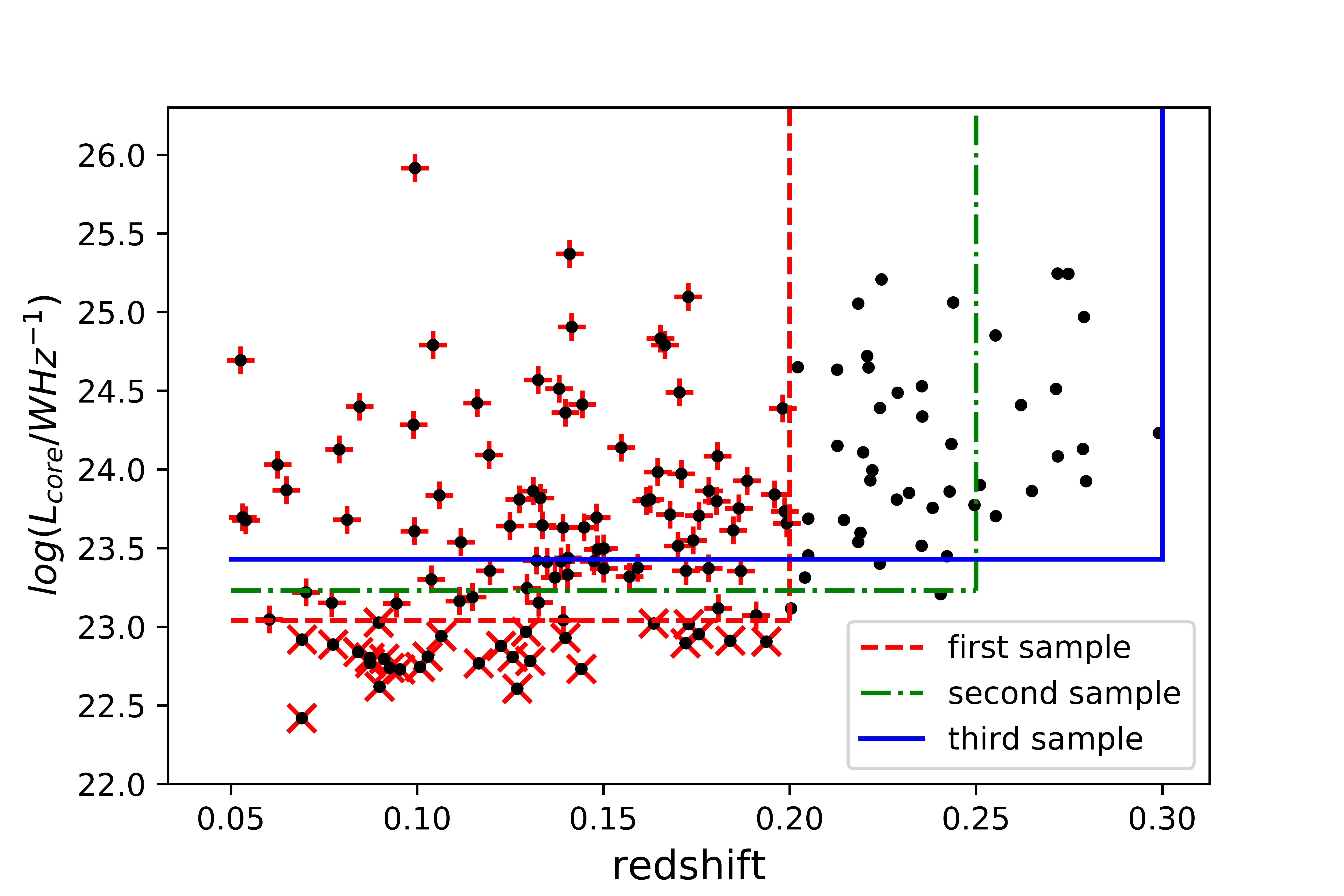}
	\caption{Radio luminosity of the core versus redshift for FRII radio galaxies used in this study. Three different subsamples are defined based on the redshift and luminosity cuts displayed by red (dashed) , green (dash-dotted) and blue (solid) lines. Core and non-core FRIIs are selected as discussed in Section \ref{sec3}. For the first subsample (red), we mark core and non-core FRIIs with the red plus and cross signs respectively. FRIIs with no radio detection at their cores are not displayed. They all lie below the luminosity cuts in this plot and are labeled as non-core FRIIs. FRII HERGs and LERGs are shown with the same symbol in this plot.}
	\label{fig2}
 \end{figure}

The core radio luminosities with respect to the redshifts for FRII radio galaxies with a radio detection at the core are presented in Fig. \ref{fig2}. The FRII galaxies with no detection at the core are not included in this plot.  In order to construct the radio complete samples of core FRIIs and the corresponding control samples of non-core FRIIs, we have defined three subsamples by applying upper cuts for redshifts. The first subsample includes galaxies with a redshift which is lower than 0.2; the second one consists of galaxies with a redshift which is lower than 0.25 and the last subsample contains galaxies with a redshift which is lower than 0.3.
\begin{table}
	\centering
	\centering
	\caption{ Number of FRII core and non-core in each class of excitation, used in this study.}
	\begin{tabular}{|c|c|c|c|c|}
		\hline
		Excitation&\multicolumn{2}{|c|}{LERGs}&\multicolumn{2}{|c|}{HERGs}\\
		\hline
		subsample&core&non-core&core&non-core\\
		\hline
		First&70&170&14&16\\
		\hline
		Second&83&240&16&21\\
		\hline
		Third&76&281&14&24\\
		\hline
	\end{tabular}
	\label{tab1}
\end{table}

 We have divided each subsample based on the core radio luminosity into two groups of core and non-core FRIIs. Three different values of radio completeness limits are defined for each subsample. The values are calculated from the redshift cuts and the completeness limit of the FIRST survey ($\sim$ 1 $mJy$). FRII radio galaxies with the core radio luminosities above these values are considered as core FRIIs. Non-core FRIIs are those with core radio luminosities below these completeness values as well as FRIIs with no detection of the core. The latter are not included in the plot. Those with no detection are well below the completeness limit as $5\sigma$ detection ($\sim$ 0.75 $mJy$) is less than 1 mJy completeness limit of FIRST. The corresponding luminosities based on 1 mJy completeness limit at each redshift limit are calculated using $L_{rad}=4\pi D_{L}^2f_{rad}/(1+z)^{1+\alpha}$ \citep{singal2011radio}, where $D_L$ is the luminosity distance and $\alpha$ is the radio spectral index which is assumed to be -0.7 in this work.

\begin{figure}[h!]	
	\begin{subfigure}{\columnwidth}
		\includegraphics[scale=0.5]{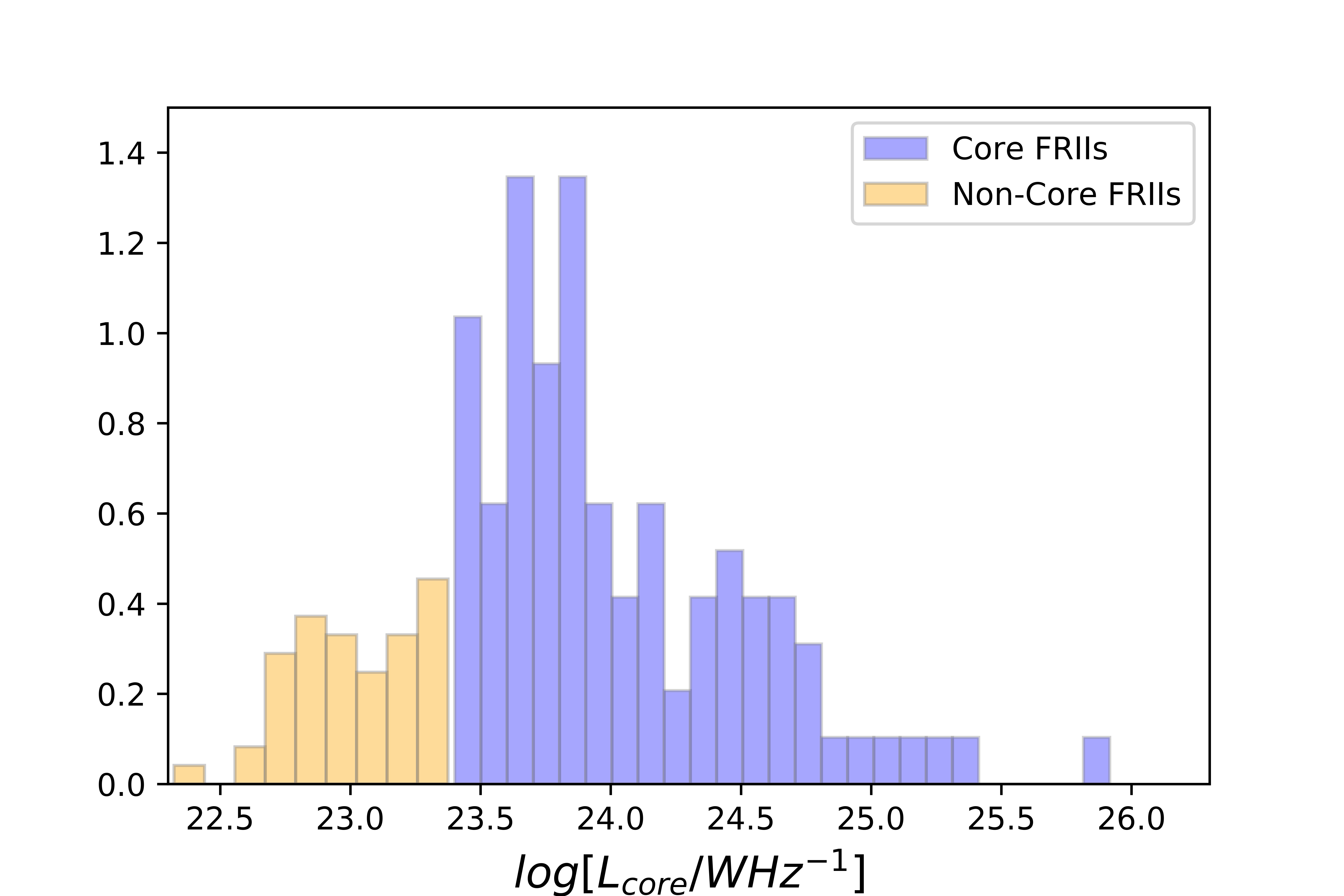}
	\end{subfigure}
	\begin{subfigure}{\columnwidth}
		\includegraphics[scale=0.5]{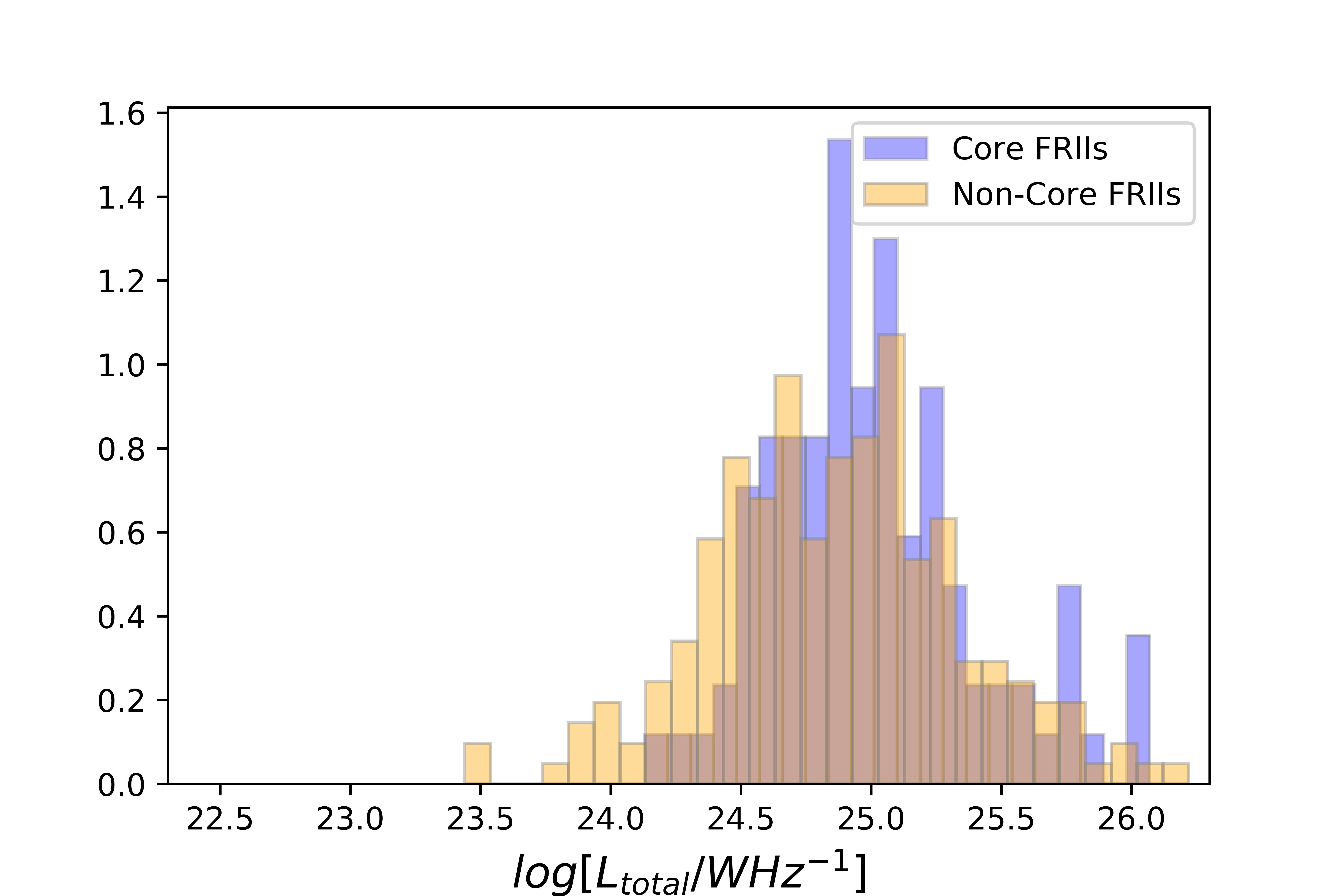}
	\end{subfigure}
\caption{The distribution of the core (top) and total (bottom) radio luminosities for the core (blue) and non-core (orange) FRII radio galaxies. The area under the histograms sum up to one. The plots include both HERG and LERG sources from the first subsample described in Section 3.} 
\label{fig3}	
\end{figure}

\begin{figure}[h!]
	\centering
	\includegraphics[width=\columnwidth]{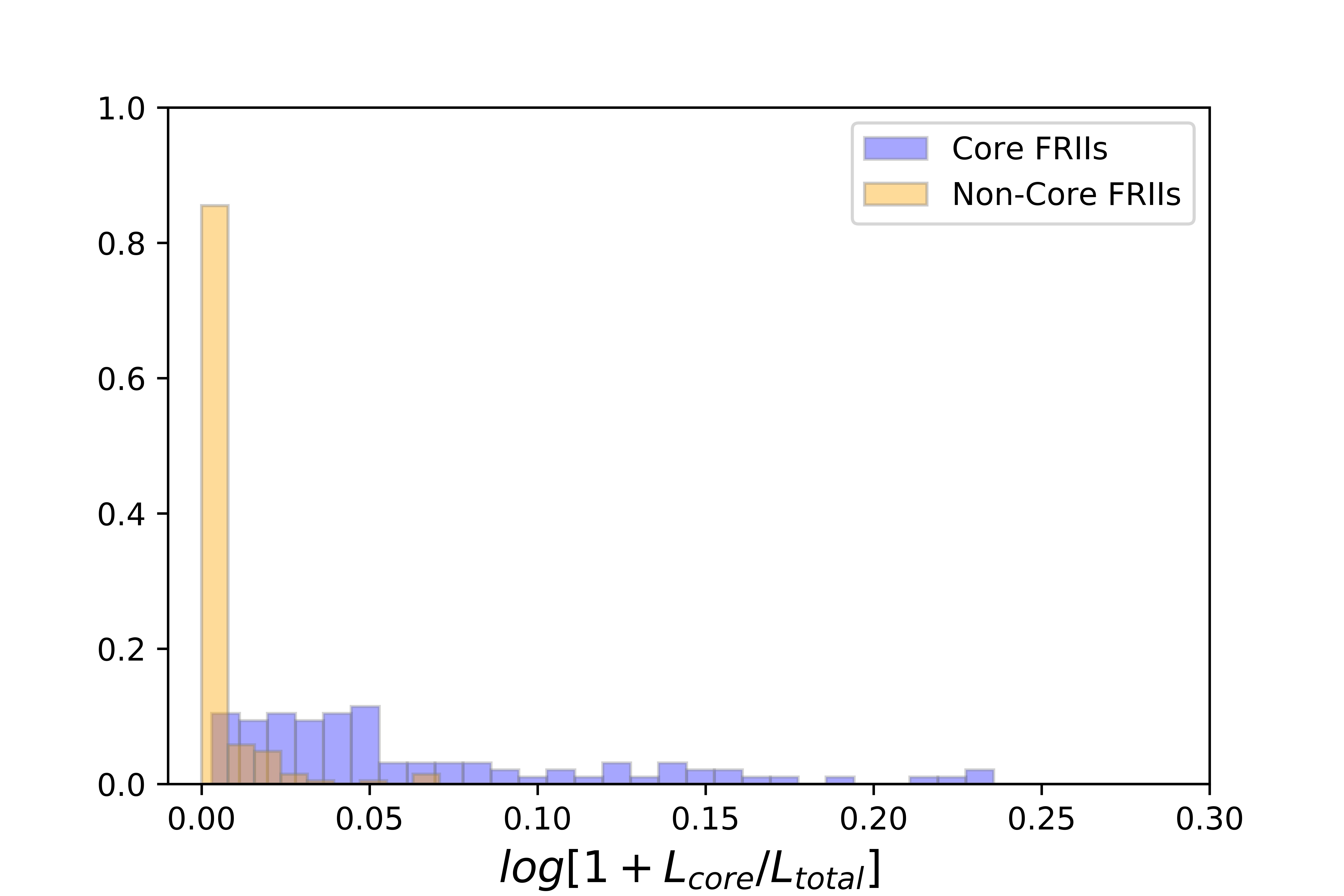}
	\caption{The ratio of the core radio luminosity to the total radio luminosity for the core (blue) and non-core (orange) FRIIs in logarithmic scale is shown. The area under the histogram sums up to one. The plot includes both HERG and LERG sources from the first subsample described in Section 3.} 
	\label{fig4}
\end{figure}

\begin{figure}[h!]
	\centering
	\includegraphics[width=\columnwidth]{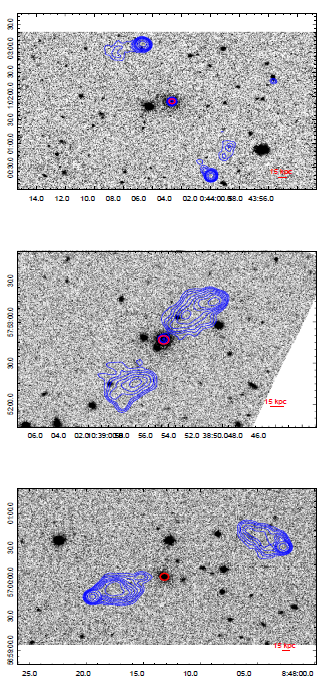}
	\caption{Three examples of FR II radio galaxies in this study. The 1.4 GHz radio contours are overlaid on the SDSS optical images at r-band. The red circles at the center of the images show the host galaxies within 3 arcseconds radius. Contour levels of 0.001, 0.0015, 0.002, 0.003, 0.004, 0.005, 0.006, 0.008, 0.016, and 0.032 Jy are shown.
		The physical scales (15 kpc) are illustrated on the bottom-right of the images.
		The top panel shows an FR II radio galaxy selected to have a core radio emission. The middle and the bottom panels show FR II radio galaxies with a very weak core and with no detection for the core respectively.}
	\label{fig5}
\end{figure}

The values are $1.1\times10^{23}\ W Hz^{-1}$, $1.7 \times 10^{23}\ W Hz^{-1}$ and $2.7 \times 10^{23}\ W Hz^{-1}$ correspond to the redshifts of 0.2, 0.25 and 0.3 respectively. Thus, e.g., FRII radio galaxies in the first subsample (z$<$ 0.2) with the core radio luminosities more than $1.1\times10^{23}\ W Hz^{-1}$ are grouped as core FRIIs and those below this threshold are grouped as non-core FRIIs. Radio galaxies without detected cores are also labeled as non-core FRIIs. The redshift and radio luminosity cuts are presented in Fig. \ref{fig2}.

The resultant subsamples are listed in Table \ref{tab1}. The table also includes the number of HERG and LERG objects in each subsample.The distributions of the total and core radio luminosities for core and non-core FRIIs are plotted in Fig. \ref{fig3}. The ratio of the core to total luminosities is also included in Fig. \ref{fig4}. Galaxies with no detection for the core have zero values in this plot. Throughout this paper, we used the first subsample to illustrate the results as the plots for the other two samples are equivalent. In Sections 4 and 5, we will show that the final results from the first subsample are in agreement with the results from the second and the third subsamples. 

Fig. \ref{fig5} shows three examples of FR II radio galaxies in this study. Top panel shows an example of radio galaxies with the core detection above the completeness limit defined in this study. The middle and bottom panels are FRIIs with a weak core or no detection of the core, respectively.
\section{The core versus non-core FRIIs: overall properties}\label{sec4} 

\begin{figure*}
	\begin{subfigure}[t]{\linewidth}
		\includegraphics[scale=0.5]{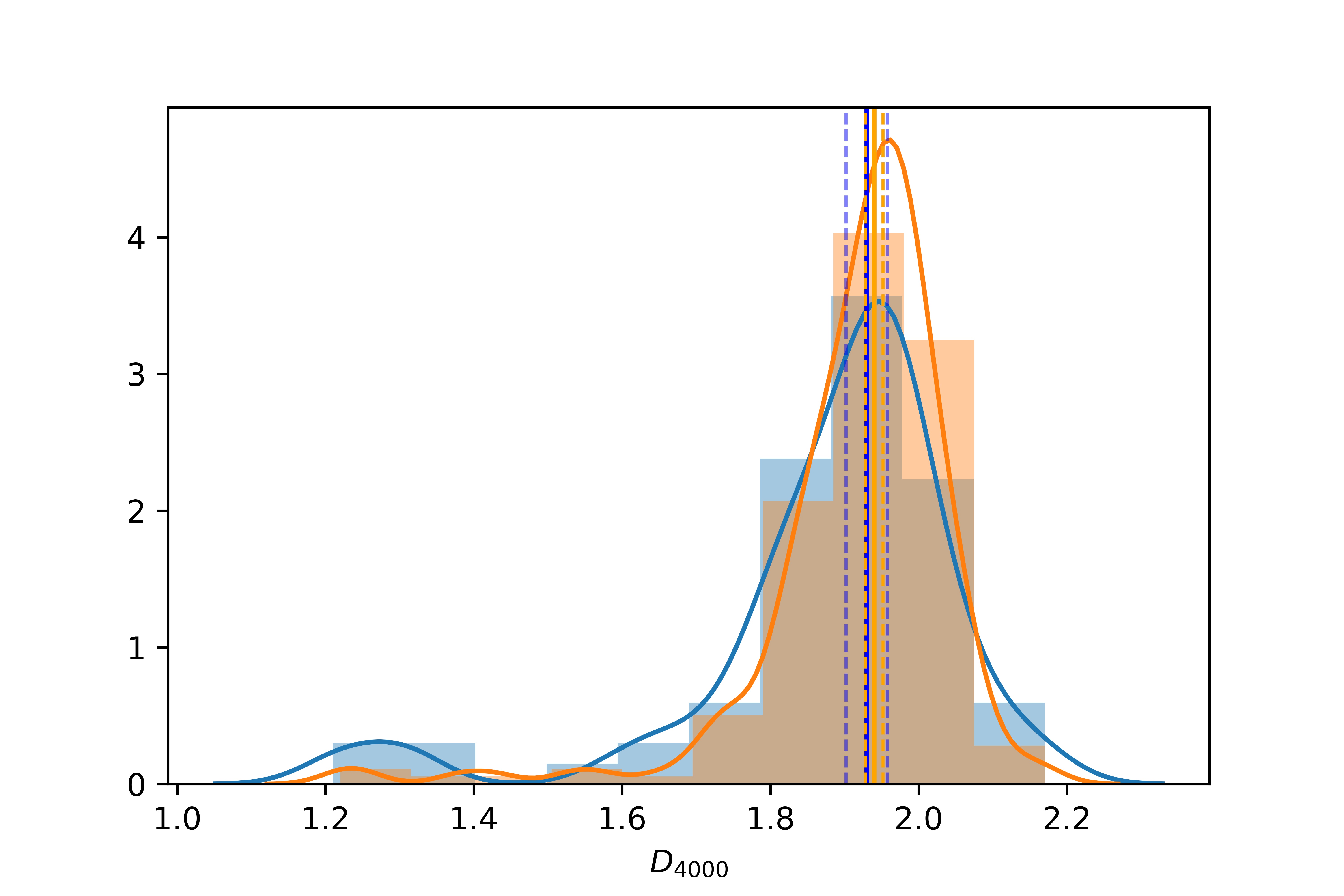}
		\includegraphics[scale=0.5]{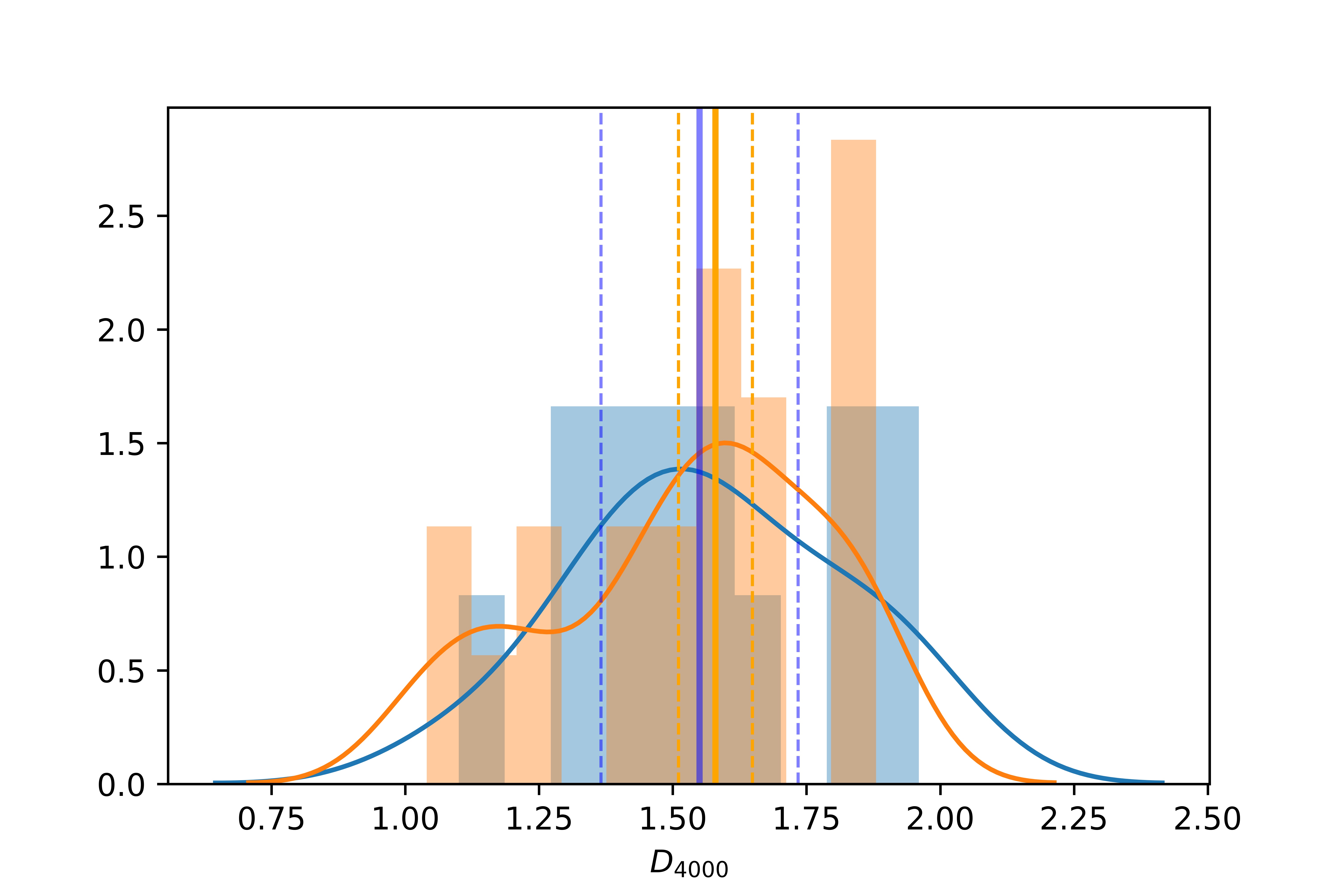}
	\end{subfigure}

	\begin{subfigure}[t]{\linewidth}
		\includegraphics[scale=0.5]{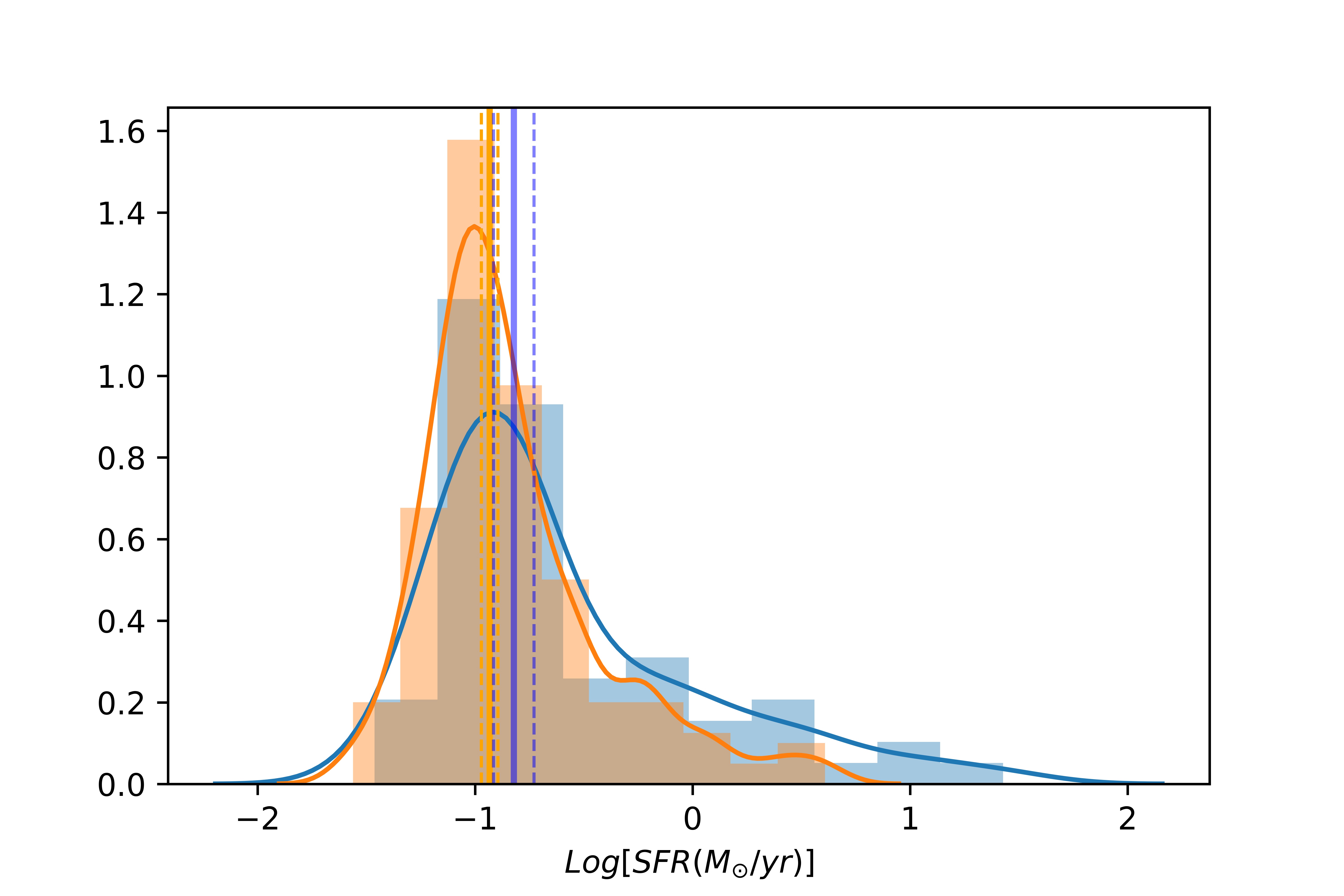}
		\includegraphics[scale=0.5]{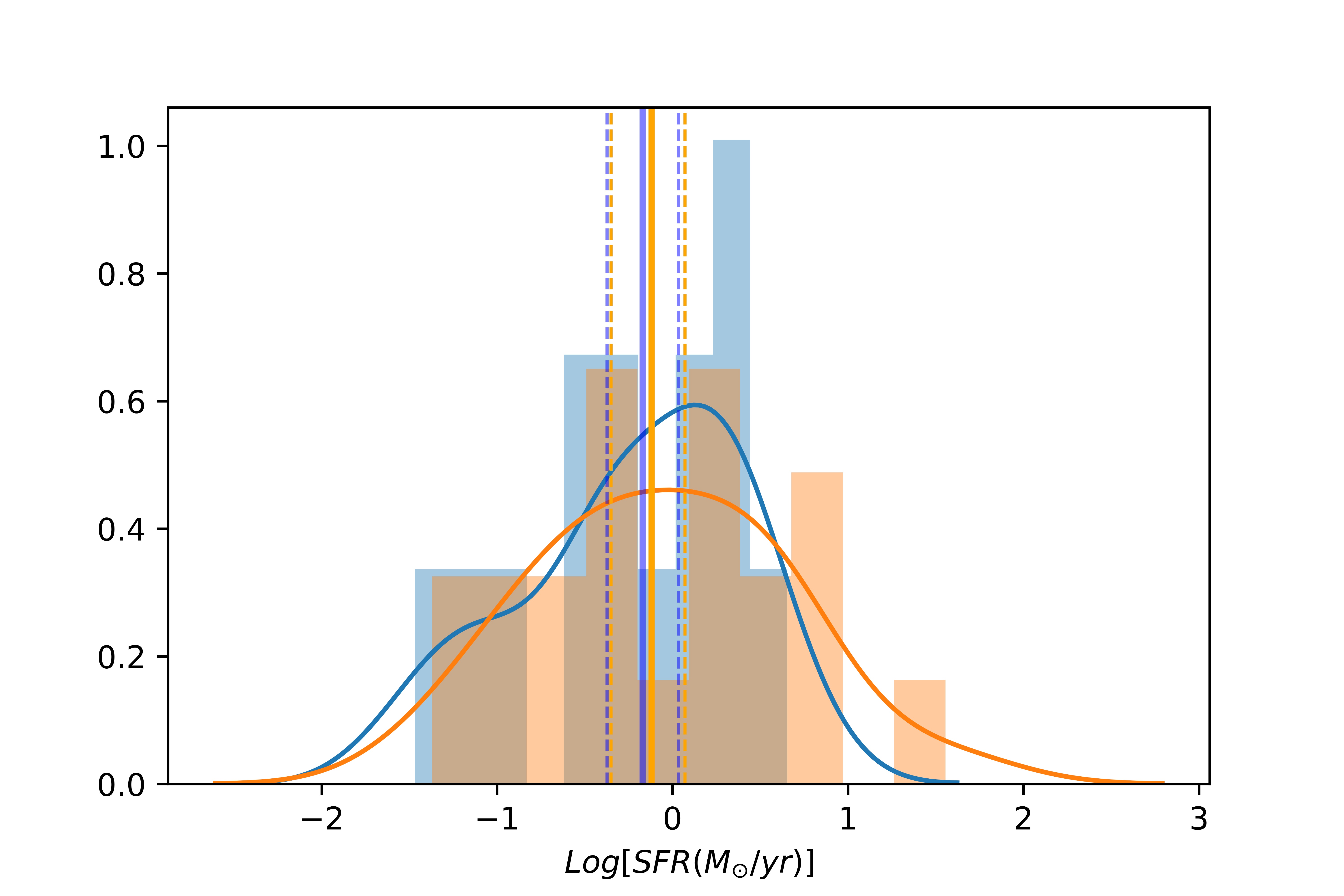}
	\end{subfigure}

	\begin{subfigure}[t]{\linewidth}
		\includegraphics[scale=0.5]{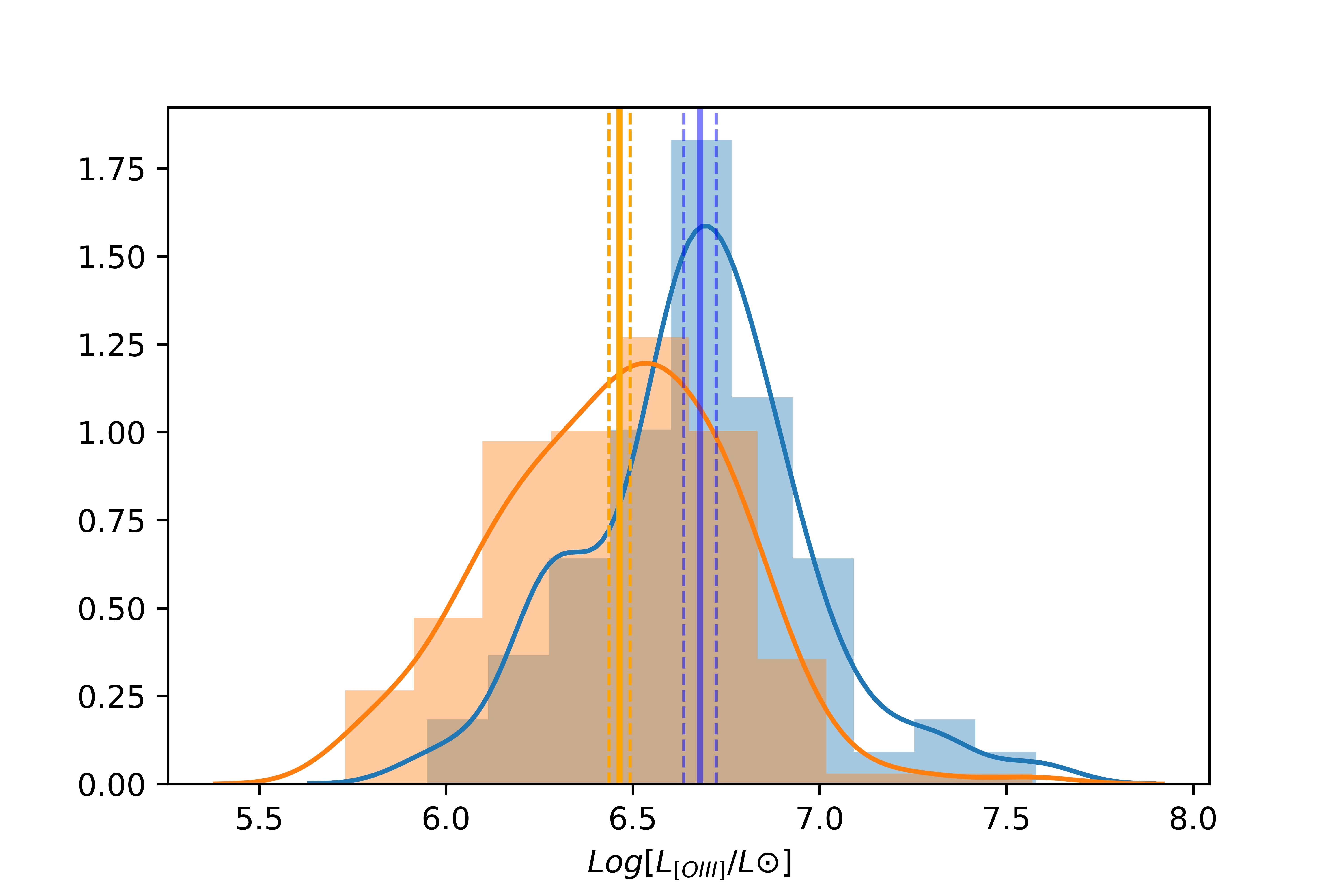}
		\includegraphics[scale=0.5]{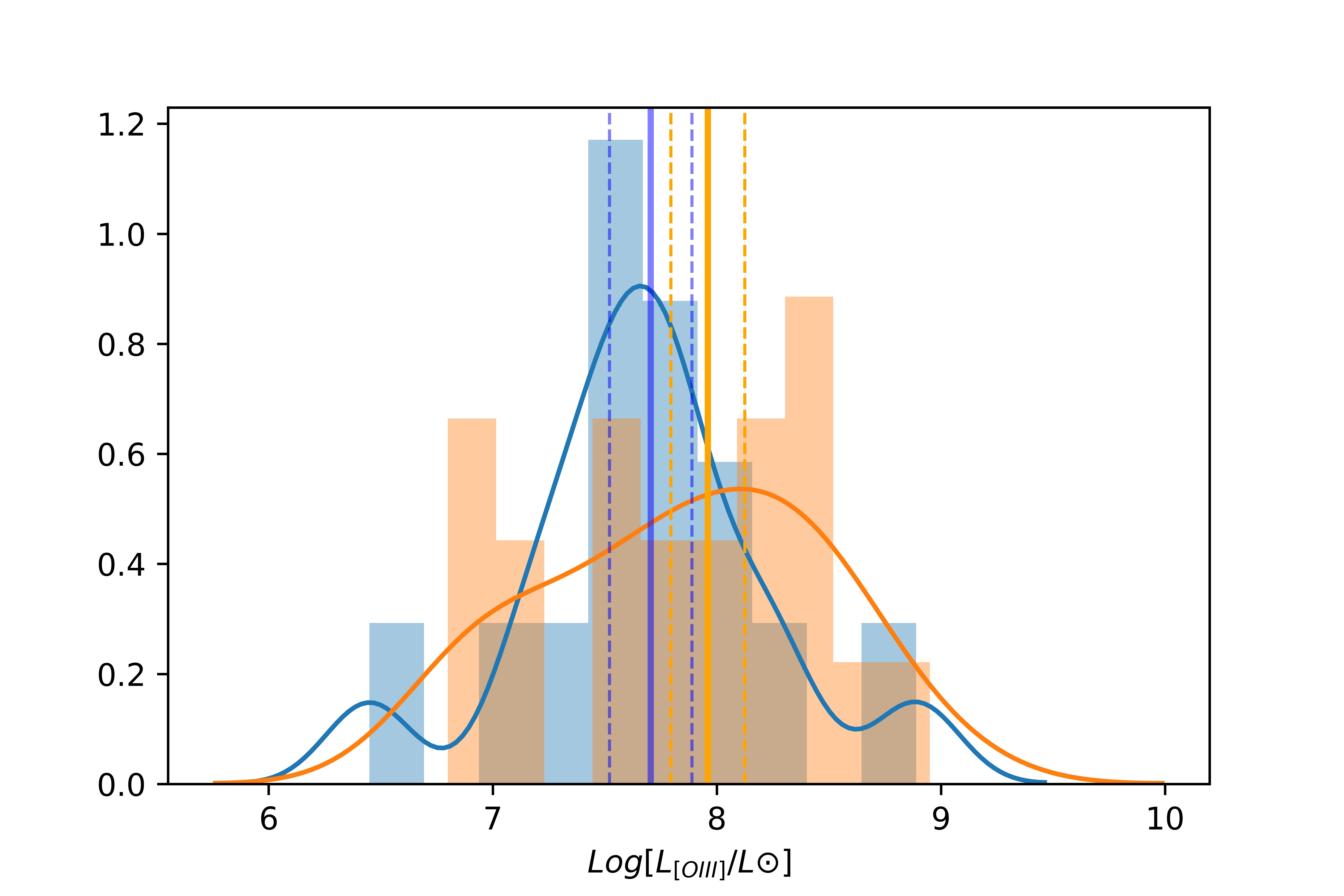}
	\end{subfigure}    
	\caption{ The distribution of 4000-\AA\ break, star formation rate and [OIII] luminosity for the first subsample introduced in Section \ref{sec3}. These diagrams show the differences between the distributions of core (blue) and non-core (orange) FRII radio galaxies. The area underneath each curve sums up to one. The vertical solid lines show the median values and the dashed lines show the upper and the lower limits of the errors for the median values. The LERGs and HERGs are presented in the left and right panels respectively. The plots show the result of non-matched samples as listed in Table 2. Only sources with detected [OIII] luminosities are considered in the lower plots.}
\label{fig6}
		
\end{figure*}

In this section, we compare the host galaxy properties of the core FRIIs with non-core FRIIs. The comparison has been done for each of the three subsamples defined in Section \ref{sec3}. The HERGs and LERGs have been considered separately for each subsample. Optical properties listed in Section \ref{sec2.2} have been investigated. Examples of  distribution histograms for $D_{4000}$, SFR and [O III] luminosity of the first subsample (z$<$ 0.2) for HERGs and LERGs are shown in Fig. \ref{fig6}. Different colours represent core (blue) and non-core (orange) FRII radio galaxies. In order to quantify the significance of the differences and to assign a probability according to the differences between the two distributions, we applied the Kolmogrov-Smirnov (K-S) test. This statistical test is based on the maximum difference between an empirical and a hypothetical cumulative distribution (one-sample K-S test) or two empirical distributions (two-sample K-S test) that are extracted from the normalized cumulative histograms of each parameter \citep{massey1951kolmogorov}. The values of the probabilities report the levels at which the null hypothesis is rejected. We report the differences above one sigma noise level which is equal to the probability of 68 percent.

\begin{table*}
	\caption{The results of the K-S test and the probabilities calculated based on the significance levels of each property. The (+)/(-) sign means this property is higher/lower for core FRIIs. The surface mass density ($\mu_{50}$), optical size ($R_{50}$), black hole mass ($M_{BH}$), [OIII] luminosity ($L_{[OIII]}/L_{\odot}$), 4000-\AA\ break ($D_{4000}$), stellar mass ($M_{\star}$), concentration ($R_{90}/R_{50}$), color ($g-r$), redshift (z), star formation rate (SFR) and specific star formation rate (sSFR), as described in Section 2.2, are listed in the table. The results are biased by the redshift as discussed in Section 4. This table presents the results for the LERG FRIIs.} 
	\small
	\centering
	\begin{tabular}{|c|c|c|c|c|c|c|c|c|c|c|c|}
		\hline
		subsample&$\mu_{50}$&$R_{50}$&$M_{BH}$&$L_{[OIII]}/L_{\odot}$&$D_{4000}$&$M_{\star}$&$R_{90}/R_{50}$&$g-r$&z&SFR&sSFR\\
		\hline
		\multirow{2}{*}{First}&\multirow{2}{*}{$<\sigma$}&\multirow{2}{*}{$<\sigma$}&\multirow{2}{*}{$<\sigma$}&99.94&\multirow{2}{*}{$<\sigma$}&\multirow{2}{*}{$<\sigma$}&\multirow{2}{*}{$<\sigma$}&82.19&87.27&84.37&85.15\\
		&&&&(+)&&&&(-)&(+)&(+)&(+)\\
		\hline
		\multirow{2}{*}{Second}&\multirow{2}{*}{$<\sigma$}&\multirow{2}{*}{$<\sigma$}&\multirow{2}{*}{$<\sigma$}&99.98&98.54&\multirow{2}{*}{$<\sigma$}&\multirow{2}{*}{$<\sigma$}&\multirow{2}{*}{$<\sigma$}&98.57&\multirow{2}{*}{$<\sigma$}&87.58\\
		&&&&(+)&(-)&&&&(+)&&(+)\\
		\hline
		\multirow{2}{*}{Third}&\multirow{2}{*}{$<\sigma$}&\multirow{2}{*}{$<\sigma$}&83.03&99.99&\multirow{2}{*}{$<\sigma$}&\multirow{2}{*}{$<\sigma$}&\multirow{2}{*}{$<\sigma$}&\multirow{2}{*}{$<\sigma$}&97.65&94.94&98.57\\
		&&&(-)&(+)&&&&&(+)&(+)&(+)\\
		\hline
	\end{tabular}\label{tab2}
\end{table*}

\begin{figure}[h!]
	\centering
	\includegraphics[width=\columnwidth]{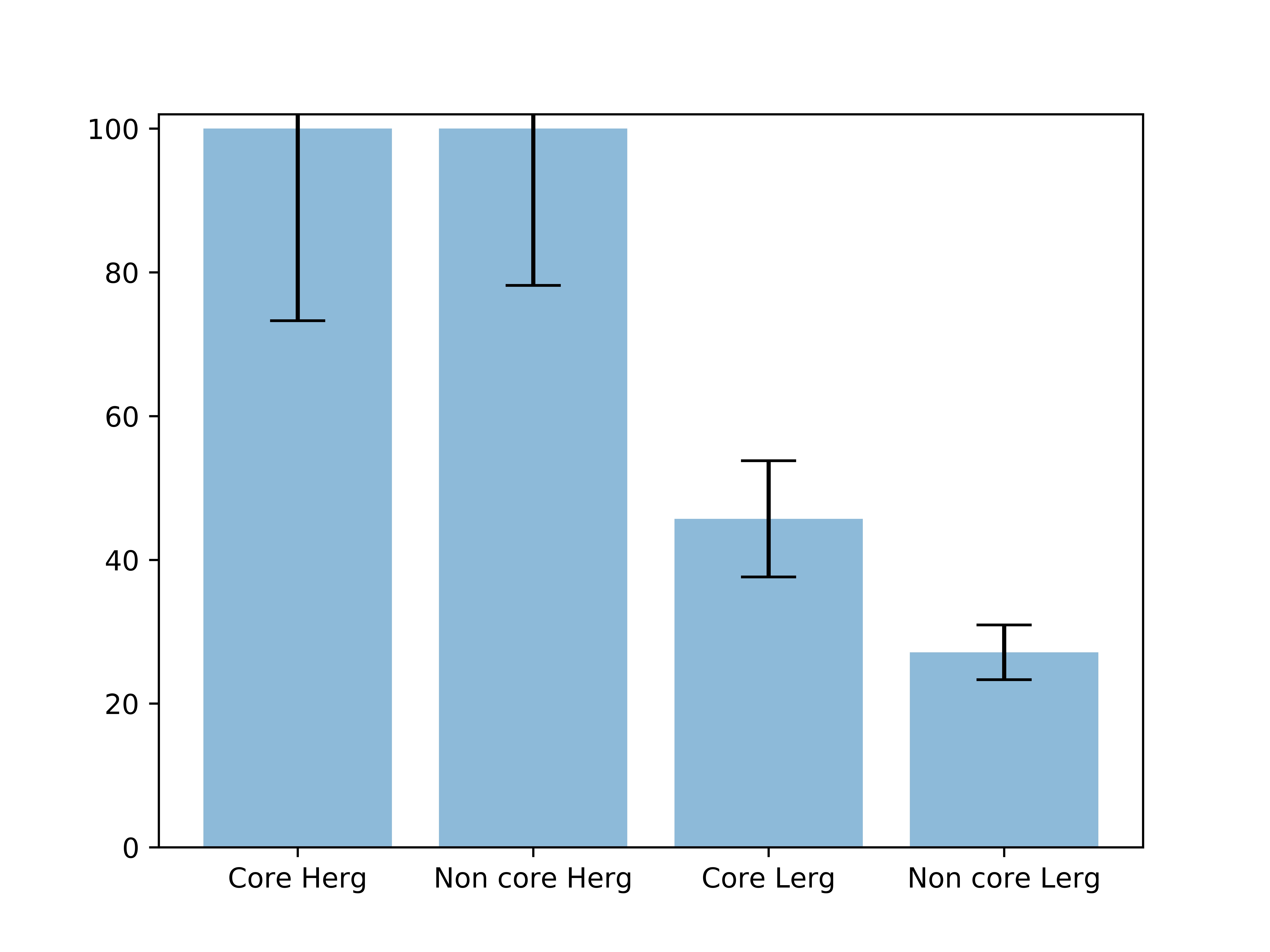}
        \caption{The fraction (percentage) of sources with [O~III] emission line detection for each type of FRIIs from the first subsample. The results for the LERGs include non-matched samples.}
	\label{figMore}
\end{figure}

To compute the probabilities associated with the significance of differences, the maximum difference of two cumulative distributions ($D=max|F_n(x)-F_m(x)|$) and the number of objects of each distribution are required \citep{young1977proof}. Here, $n$ and $m$ are the number of objects in core and non-core distributions. The estimated probabilities of each physical parameter for each subsample for LERGs are shown in Table~\ref{tab2}.  The percentages greater than 99.7 indicate that the significance of differences in the two distributions are above $3 \sigma$. The values more than 95 and 68 percents, determine the significance levels more than 2~$\sigma$ and 1~$\sigma$ respectively.
None of the differences between core and non-core FRIIs in the HERG sample are significant due to the small sample size as illustrated in Fig. \ref{fig6} so we do not report them in the table.

For [O III] luminosities, we have used two methods. In the first method, we keep the luminosities with $S/N \geq 2.5$ and use L$_{O III,min}$ for the luminosities with $S/N < 2.5$. Then the distributions are plotted and the probabilities are calculated. In the second method, we keep the luminosities with $S/N \geq 2.5$ and use L$_{O III,max}$ for the luminosities with $S/N < 2.5$. The results out of the two methods are analogous and thus in Table~\ref{tab2}, we just have reported the result of the second method. Fig.7 shows the fraction of sources with [O III] emission line detection ($S/N \geq 2.5$) in the first subsample and for each of core/non-core and HERG/LERG samples. 

There are some differences above 1~$\sigma$ in the first set for LERGs. This includes [O III] luminosity, g-r colour, SFR and sSFR. The distributions of the redshifts are also different. Core FRIIs have on average higher [O III] luminosities which is confirmed with more than 3 $\sigma$ significance level. The host galaxies in core FRIIs are slightly bluer (lower g-r) and have higher SFRs and sSFRs. Some more significant differeces are found for the second and the third sets. These include redshifts which may produce biases to the subsamples. There are no significant differences in the stellar masses in all three subsamples so the results are not biased by the stellar mass. Note that the stellar mass has the most important role to trigger radio-mode AGN activity in galaxies \citep{sabater2019lotss}. In the following section, we will remove the bias from the redsift to get more robust results. The results for [O III] luminosity are described and investigated in more detail in Section \ref{sec6}.  

\begin{table*}
	\caption{The results of the K-S test and the associated probabilities for each parameter calculated for the core and non-core FRIIs in the same stellar mass-redshift bins. The (+)/(-) sign means this property is higher/lower for core FRIIs. The surface mass density ($\mu_{50}$), optical size ($R_{50}$), black hole mass ($M_{BH}$), [OIII] luminosity ($L_{[OIII]}/L_{\odot}$), 4000-\AA\ break ($D_{4000}$),  concentration ($R_{90}/R_{50}$), color ($g-r$), star formation rate (SFR), as described in Section 2.2, are listed in the table. The only significant difference between the two samples is for [O III] luminosity. This table presents the results for the LERG FRIIs.}
	\begin{center}
		\begin{tabular}{|c|c|c|c|c|c|c|c|c|}
			\hline
			subsample&$\mu_{50}$&$R_{50}$&$M_{BH}$&$L_{[OIII]}/L_{\odot}$&$D_{4000}$&$R_{90}/R_{50}$&$g-r$&SFR\\
			\hline
			\multirow{2}{*}{First}&\multirow{2}{*}{$<\sigma$}&\multirow{2}{*}{$<\sigma$}&\multirow{2}{*}{$<\sigma$}&97& \multirow{2}{*}{$<\sigma$}&\multirow{2}{*}{$<\sigma$}&\multirow{2}{*}{$<\sigma$}&\multirow{2}{*}{$<\sigma$}\\
			&&&&(+)&&&&\\
			\hline
			\multirow{2}{*}{Second}&\multirow{2}{*}{$<\sigma$}&\multirow{2}{*}{$<\sigma$}&\multirow{2}{*}{$<\sigma$}&99& \multirow{2}{*}{$<\sigma$}&\multirow{2}{*}{$<\sigma$}&\multirow{2}{*}{$<\sigma$}&\multirow{2}{*}{$<\sigma$}\\
			&&&&(+)&&&&\\
			\hline
			\multirow{2}{*}{Third}&\multirow{2}{*}{$<\sigma$}&\multirow{2}{*}{$<\sigma$}&\multirow{2}{*}{$<\sigma$}&90& \multirow{2}{*}{$<\sigma$}&\multirow{2}{*}{$<\sigma$}&\multirow{2}{*}{$<\sigma$}&\multirow{2}{*}{$<\sigma$}\\
			&&&&(+)&&&&\\
			\hline
		\end{tabular}\label{tab3}
	\end{center}	
\end{table*}

 \section{The core versus non-core FRIIs: 2D matched analysis}\label{sec5}
As discussed in Section \ref{sec4}, the second and the third subsamples from the LERG type show significant differences between core and non-core FRII galaxies (Table~\ref{tab2}). In this section, we intend to consider whether differences in the redshift distributions of the two samples may cause these results.  We also ensure that the results are not biased by differences in the stellar masses.

\textbf{Redshift:}
As presented in Section \ref{sec4}, core FRIIs have slightly higher redshifts than non-core FRIIs. The difference is especially significant for the second and third subsamples for which other significant differences in galaxy properties have also been reported. In order to remove any differences caused by the redshift bias (either as a direct result of galaxy evolution or a selection bias), radio galaxies with the same redshift distributions should be investigated. In the following, matched samples of core and non-core FRIIs within a redshift bin $\Delta$z=0.02 have been constructed.     

\textbf{Stellar mass:}
There is a strong correlation between the stellar mass of galaxies and the probability of galaxies hosting radio-loud AGN \citep{best2005host}. Therefore, for further consideration on the origin of radio emission in galaxies, we should fully take this into account. The stellar mass distributions of the two types of radio galaxies in this study were the same as reported in Table~\ref{tab2}. The differences in the stellar masses for three subsamples were below one sigma significance so the results were not biased by the stellar mass. Making matched samples in redshift may change the distributions of the stellar masses thus in the following, the radio galaxies in each type are selected to be matched within a stellar mass bin $\Delta$Mass=0.2. 

Based on the discussions above, three subsamples of radio galaxies from each type (core and non-core FRIIs) have been constructed. These three subsamples correspond to three redshift-radio luminosity cuts described in Section \ref{sec3}. Each subsample is composed of matched samples in the stellar mass and redshift from core and non-core FRIIs. The results of comparison between core and non-core FRIIs constructed by this method for each galaxy property are summarized in Table~\ref{tab3} as described in Section \ref{sec4}. We repeated the analysis a hundred times and reported the averaged values in Table~\ref{tab3}. One hundred matched subsamples were constructed by random selections of pairs of radio galaxies (a core FRII matched with a non-core FRII) from the same stellar mass-redshift bins.  
As in Section \ref{sec4}, we only reported the results for the LERGs as the small sample sizes of HERGs do not allow us to investigate 2D matched comparisons.

Table~\ref{tab3} shows that the results for [O III] luminosity are consistent with the results in Section \ref{sec3}. Other differences are not replicated by the 2D analysis. This indicates that they may be caused by redshift bias. Since the core and non-core FRII samples are matched in the stellar mass and redshifts, these properties are not listed in Table~\ref{tab3}. The sSFR is also removed. It presents the same result as SFR when both samples are matched in the stellar masses. The difference in [O III] luminosity is significant with the confidence levels of 97, 99 and 90 percent for the first, second and the third subsamples respectively. Therefore, core FRIIs have higher [O III] luminosity than non-core FRIIs. The origin of this difference can be the presence of ionized gas due to star formation activity or AGN activity.
Since in the 2D analysis both samples show the same distributions of SFR, colour and $D_{4000}$, we relate the excess [O III] emission in core FRIIs to AGN activity rather than star formation activity.  We discuss the results further in the next section. 

\section{The origin of [O III] luminosity}
\label{sec6}

\begin{figure*}
		\begin{subfigure}{\linewidth}
		   \includegraphics[scale=0.5]{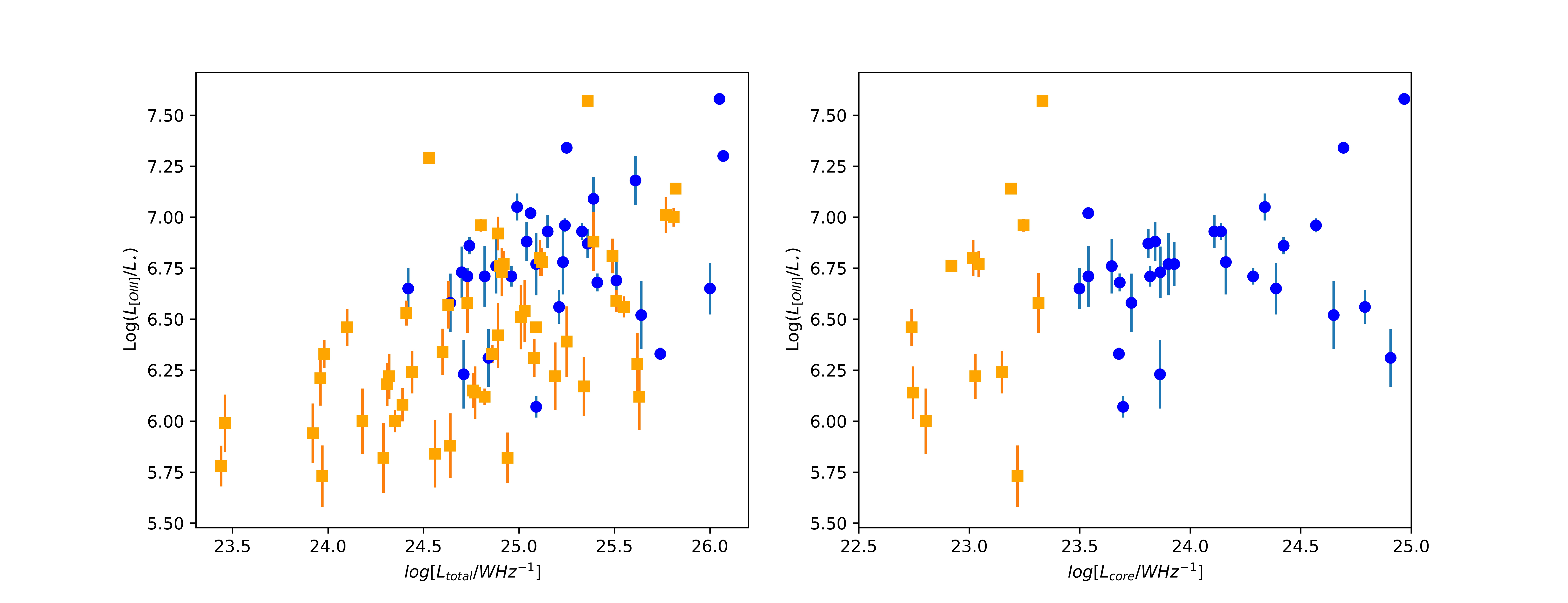}
		\end{subfigure}
		\begin{subfigure}{\linewidth}
			\includegraphics[scale=0.5]{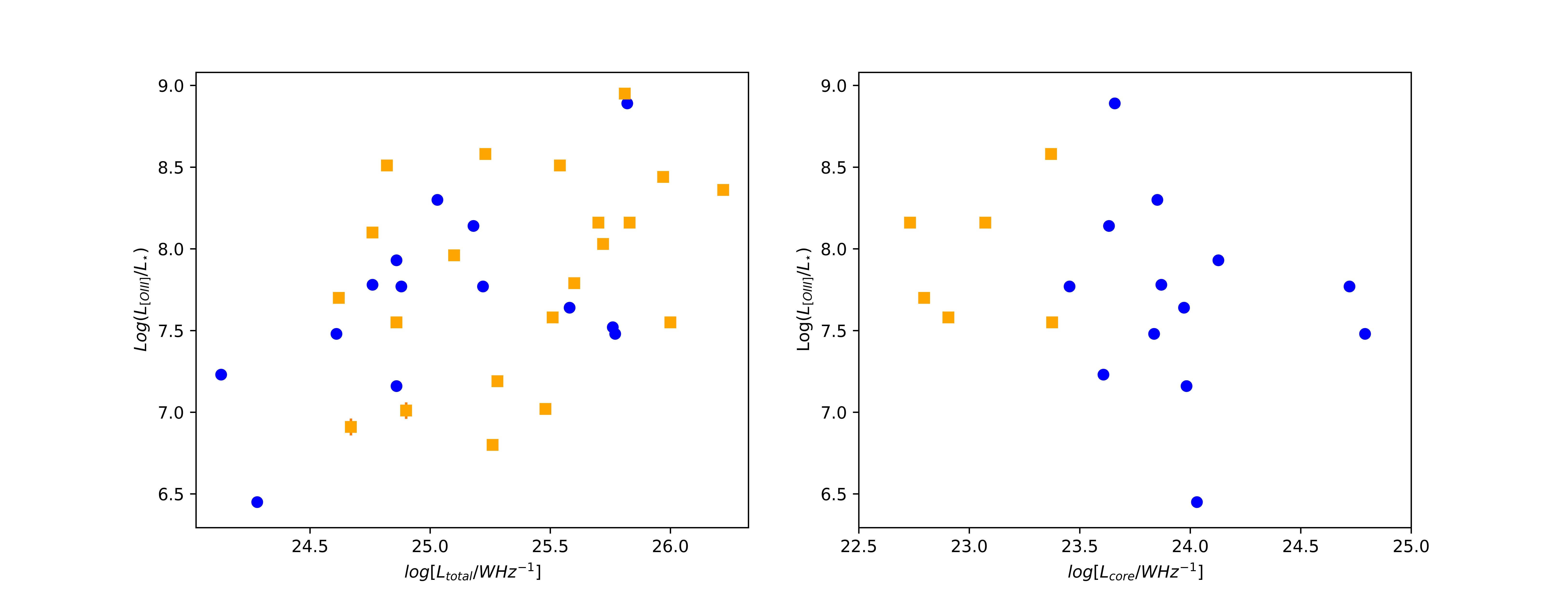}
		\end{subfigure}
		\caption{ The [O III] luminosities versus total (left-hand panel) and core (right-hand panel) radio luminosities.The upper panels show the results for LERGs while the lower panels present the HERGs. The error bars indicate the noise level. Blue circles represent core FRIIs and orange squares represent non-core FRIIs. The data includes non-matched sources from the first subsample. Only sources with the detected [O III] line and radio luminosities are plotted.} 
		\label{fig8}
\end{figure*}
The relation between [O III] luminosity and the total and core radio luminosities of radio galaxies have been investigated in previous studies. In this regard, \citet{buttiglione2010optical} show that there is a correlation between [OIII] luminosity with either core or total radio luminosity for a sample of powerful FR radio galaxies. Using a larger sample, \citet{capetti2017fricat} showed a large spread in both total radio luminosities and [O III] line luminosities for FRI radio galaxies. Similar results have also been obtained by different authors \citep{hardcastle2009active, mingo2014x}. The radio luminosities of the core have been investigated by \citet{baldi2015pilot,baldi2019high} who have shown that FRI and FR0 radio galaxies lie in the same correlation line in the radio-[O III] plane using the bolometric radio luminosity of the core versus [O III] line luminosity. They interpret this as evidence for the similarity between the nuclei of FR0s and FRIs. In terms of FRII radio galaxies, a study by \citet{capetti2017friicat} shows that there is no correlation between total radio luminosity and [O III] line luminosities. Fig. \ref{fig8} shows [O III] line luminosities versus total (left-hand panels) and core (right-hand panels) radio luminosities for FRIIs in the first subsample (z$<$ 0.2). The upper panels show the results for LERGs while lower panels present the HERGs. Only sources with detected [O III] line and radio luminosities are plotted. Core and non-core FRIIs are shown in blue circles and orange squares, respectively.

There is a weak correlation between the total radio luminosity and [O III] line luminosity in either HERG or LERG. The inconsistency between our results and those of \citet{capetti2017friicat} may be due to the slightly wider redshift range (z$<$ 0.2) we used compared to that of their sample (z$<$ 0.15). In terms of the core radio luminosity, a weak correlation is found for the LERGs consistent with the result obtained in Sections \ref{sec4} and \ref{sec5} while an anti-correlation is shown in HERGs. We note that, the result for HERG may be influenced by the obscuration from the dusty structure. Multiwavelength observations are needed to get a more reliable explanation for this. If we assume the result for HERG is not affected by obscuration or small sample size, then these results may have origin at different nature of HERGs as high accretion rate objects compared to the LERGs with lower accretion rates. Later in this section, we will also show that they display different behaviour in the [O III] luminosity-$D_{4000}$ plane.

\begin{figure}[h!]
	
\begin{subfigure}{\textwidth}
	\includegraphics[scale=0.6]{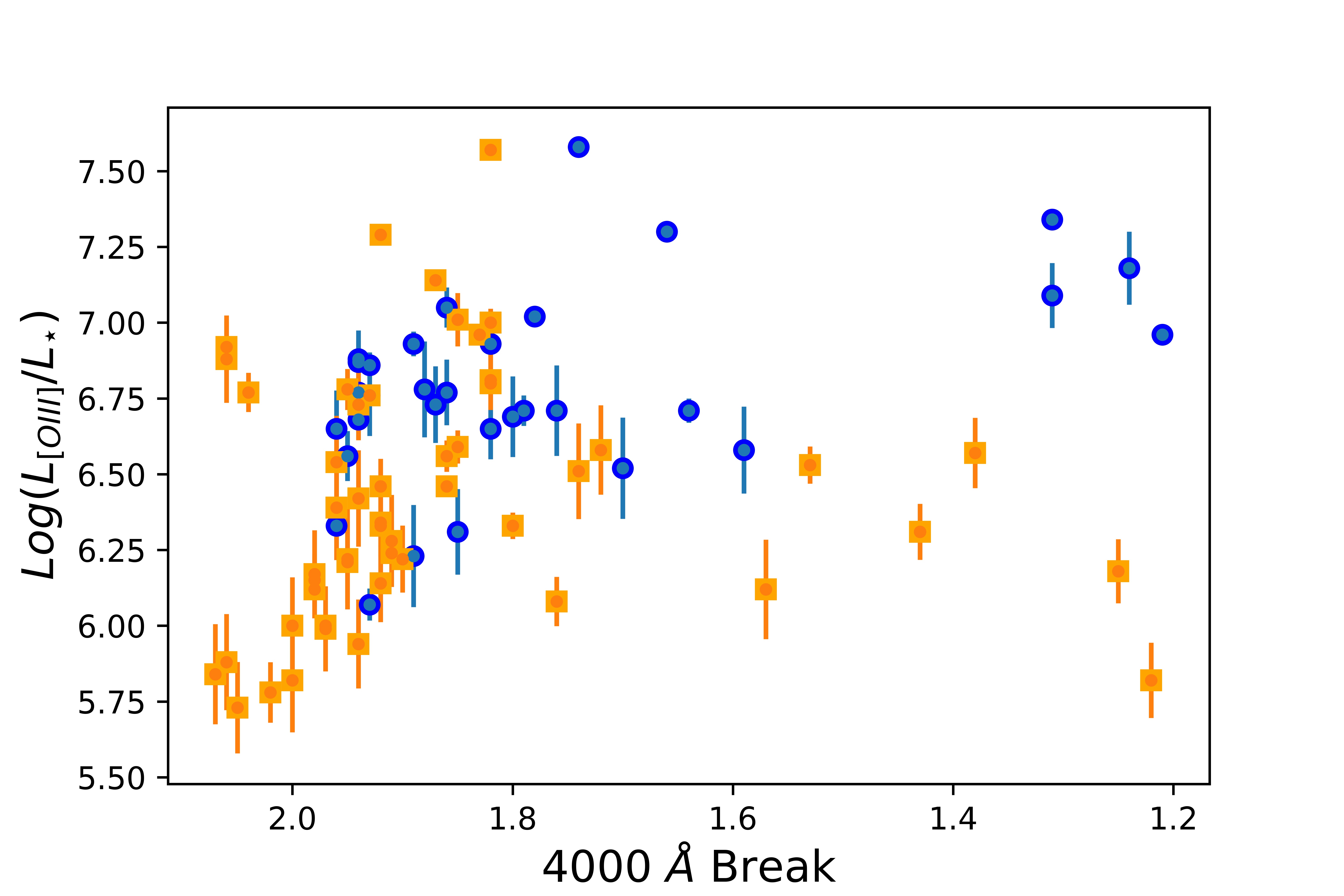}
\end{subfigure}
\begin{subfigure}{\textwidth}
	\includegraphics[scale=0.6]{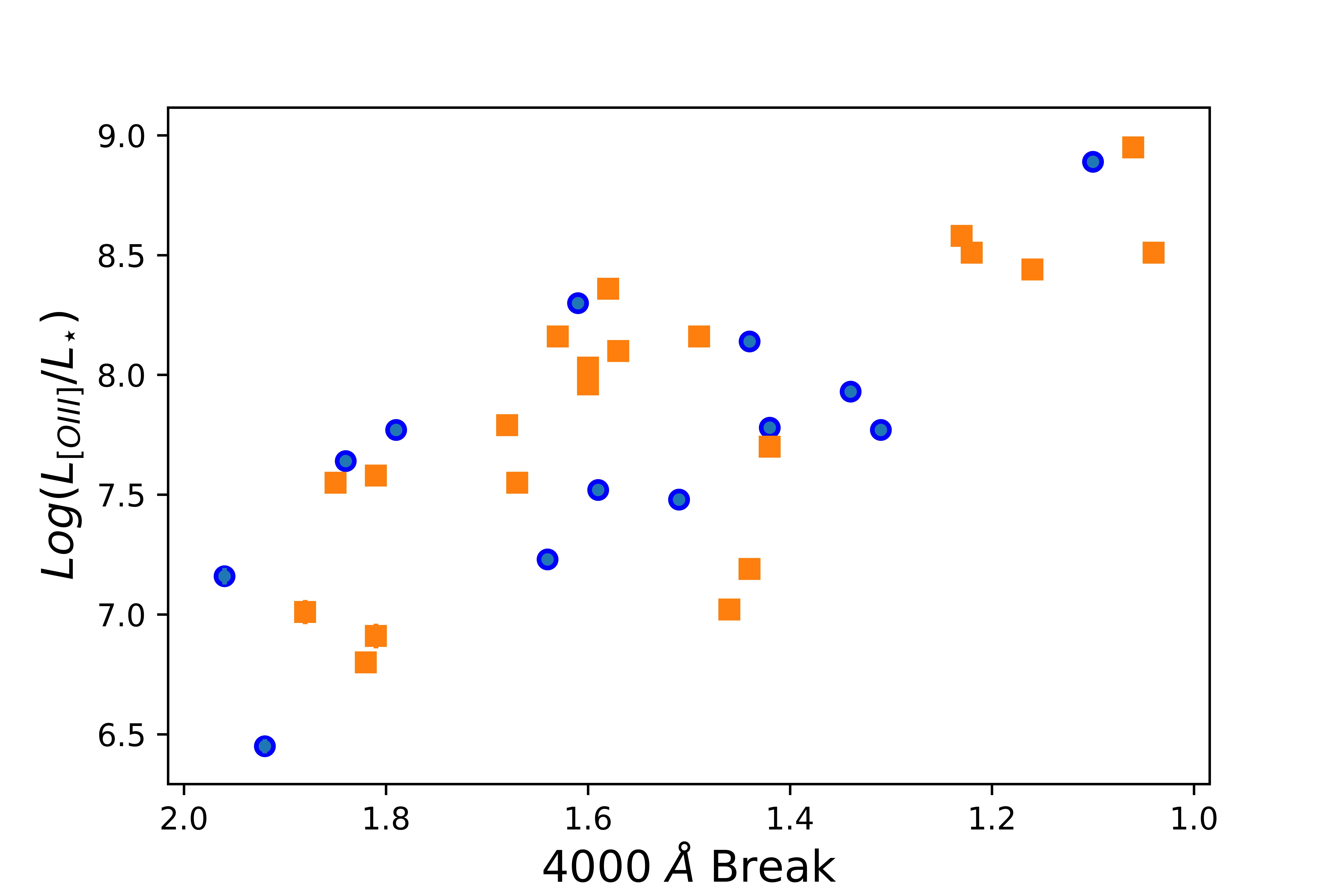}
\end{subfigure}

		\caption{ The [O III] luminosity versus 4000-\AA\ break. Blue circles represent core FRIIs and orange squares represent non-core FRIIs. The upper panel shows the result for LERGs while the lower panel presents the HERGs. The error bars indicate the noise level. The data includes non-matched sources from the first subsample. Only sources with detected [O III] line and radio luminosities are plotted.}	
\label{fig9}
\end{figure}

To investigate the origin of the relation between radio luminosity of the core and the [O III] emission line, [O III] luminosity versus $D_{4000}$ for the core and non-core FRIIs are plotted in Fig. \ref{fig9}.  The plot for LERGs shows that [O III] luminosities in core FRIIs are higher than those for non-core FRIIs in either $D_{4000}$$>$1.7 (for the early-type galaxies with older stellar populations) or $D_{4000}$$<$1.7 (for the late-type galaxies with younger stellar populations). This implies that the difference is not caused by the star formation of the host galaxies and the AGN has a large contribution to line luminosities in core FRIIs.

There are two possible scenarios to explain the results for LERGs: i) ionization of the gas which is present at the vicinity of active nucleus by the AGN light may have a large contribution to a higher [O III] luminosity detected in the core FRIIs than non-core FRIIs. This gas can also feed the SMBH to produce small scale radio jets. The lower amount or lack of this gas in non-core FRIIs leads to a lower [O III] luminosity in these galaxies. ii) the interaction of the radio jet with ambient gas results in a higher level of ionisation for the core FRIIs compared to the non-core FRIIs. In this regard, investigation of compact steep spectrum sources reveals the existence of extended emission lines aligned with the radio jets \citep{privon2008wfpc2}. \citet{tadhunter2000large} discussed the gas can be ionized either by direct interaction with the radio components, or by the diffuse photoionizing radiation fields produced in the shocks generated in such interactions. In addition, the presence of jet-driven outflow is reported at various wavelengths including the optical spectrum (\citet{o2021compact} and references therein). These works generally support our result for detection of higher [O III] luminosity in the core FRIIs compared to the non-core FRIIs.

The plot for HERGs shows no significant difference between the core and non-core FRIIs (Fig. \ref{fig9}). In addition, the two properties are correlated in HERGs for both core and non-core samples. The different behaviour of HERGs and LERGs again can have origin at the different nature of the triggering mechanism. We note that, HERGs have significantly higher [O III] luminosity (up to two orders of magnitude) compared to the LERGs. Therefore, it is also possible that the differences observed for the LERGs are not dominant in these objects. 
To obtain a conclusion for HERGs, a larger sample size is needed. Moreover, a careful consideration for the effect of obscuration on the emission line luminosities is essential.

\section{Orientation and relativistic beaming effects}

The characteristics of radio jets depend primarily on their axis orientation relative to the observer. This is the result of an anisotropic nature of radio emissions in jets. A line-of-sight close to the source jet-axis offers a view of the compact, Doppler-boosted, flat-spectrum base of the approaching jet. Therefore, orientation can influence the observed properties such as morphology \citep{harwood2020unveiling}, observed luminosities, continuum spectrum and so the spectral index of the radio jets. This effect is more pronounced in studies of compact sources compared to the extended sources as the extended sources are mostly lobe-dominated, and in studies of flat spectrum sources compared to the steep spectrum sources due to relativistic beaming. In addition, by observing at low frequencies in which the extended sources are more abundant in the radio sky, the beaming effect is decreased. In contrast, high frequency radio surveys are dominated by the compact flat spectrum sources and are largely influenced by the beaming effects \citep{de2010radio}. While a complete correction for this effect needs high resolution, multiwavelength radio observations over different periods of time, here we argue how much this could affect our results in this study.

A large fraction of beamed objects at high radio luminosities are type I AGN. The jet axis is close to the line of sight for these objects resulting in a boost in radio luminosity at relativistic speeds. These are excluded from the analysis as described in Section \ref{sec2.3}. In a similar study, Sadler et al. (2013) used four different methods to find out the fraction of beamed sources in their sample. The methods were based on the estimations for variability, the radio spetral index, the non-thermal contribution to the optical spectrum and cross-matching the catalogue with well-known beamed objects. They evaluated a lower/upper limit of 6-35 percent for the flat spectrum sources to be affected by relativistic beaming. We note that a small fraction of their sample includes broad emission line AGN while we excluded these sources from our sample.

\begin{figure}[h!]
	\begin{center}
		\includegraphics[width=\columnwidth]{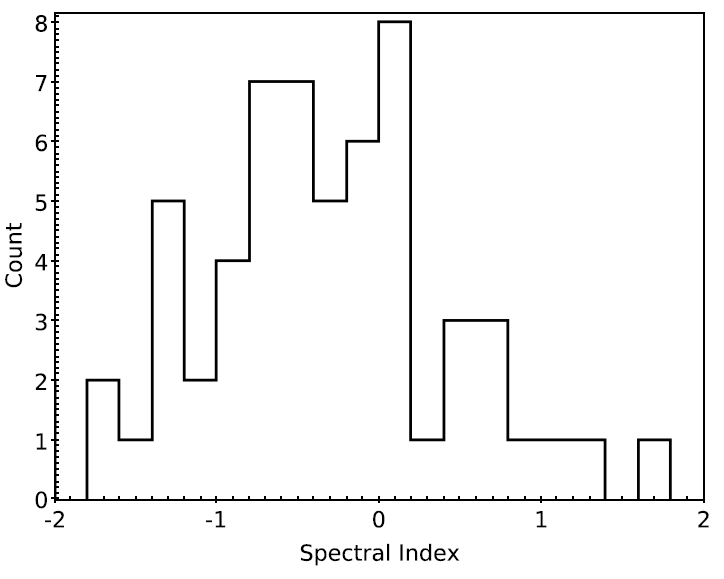}
\caption{The distribution of the spectral index for the core FRIIs of the LERG type from the first subsample calculated using the flux densities of the core component between 1.4-3 GHz. }
\label{fig10}
	\end{center}
\end{figure}

In order to investigate this for our sample, we used 3GHz VLASS data to estimate the spectral index of the core for the core FR IIs. The distribution of the spectral index for the first subsample is shown in Fig. \ref{fig10}. A few sources show no detection at 3GHz. Considering a 5 $\sigma$ noise level of 0.6 mJy ($\sigma\approx0.12$ mJy/beam) for VLASS and the 1mJy completeness cut that we applied to the FIRST sources, they all have $\alpha<$-0.6. Therefore, about 50 percent of sources show to have flat spectrum ($\alpha>$-0.5). If we assumed that one third of these sources are beamed as estimated by \citet{sadler2013local}, then less then 20 percent of our sample are affected by relativistic beaming.

This is also consistent with the results of \citet{marcha2005unification}, who argued that the effect of relativistic beaming is minimal for observed core luminosities. They reported a higher emission line luminosity in a sample of core-dominated sources compared to the FRIs. Moreover, using very-long-baseline interferometry (VLBI) observations, \citet{cheng2018parsec} showed that 4 out of 14 compact radio sources in their study (28 percent) are beamed consistent with Sadler et al (2013). Although their sample selection was different from that of our work, this could provide an estimate for the beamed sources.

\section{ Restarting sources  }

A very interesting type of FRII radio galaxies is Double-Double (DD) FRIIs \citep{mahatma2019lotss}. DD FRIIs consist of two radio lobes at each side display two distinct cycles of AGN activity.  They are also the largest radio galaxies in size. \citet{bruni2018probing} showed 80 percent of a sample of giant radio galaxies show signs of restarting radio AGN activity consistent with this picture. In addition, \citet{mingo2019revisiting} presented a new emerging type of radio galaxies called core-D which are core-dominated FRIs with the bright ends like FRIIs. They discussed how restating AGN activity may produce this type of radio structures. The radio emission from the past cycle of AGN activity (the more distant lobe) is dimmer due to the spectral aging \citep{feretti2008clusters} compared to the new cycle of that. The distant lobe is even faded away, observed as ghost cavity in X-ray images of radio galaxies \citep{jetha2008nature}.

In this section, we first argue how much our sample is contaminated by the restarting sources and whether the duty cycle of AGN activity has a large contribution in the radio emission of the core. Then, we develop an analysis to correct the possible effects of restarting activity.

Miraghaei and Best (2017) report 5 DD FRIIs in a sample of 646 FRII radio galaxies. \citet{mingo2019revisiting} report 19 DD FRIIs out of 546 FRIIs identified in the first LOFAR Two-Metre Sky Survey (LoTSS) data release. Both studies confirm a very small contribution of DD FRIIs in the total FRII samples. Thanks to the high sensitivity and resolution at MHz observations of radio galaxies by LOFAR,  \citet{mingo2019revisiting} identified 99 core-D radio galaxies which indicates a larger fraction ($\sim$20 percent) for restarting AGN activity. These sources are not recognized by \citet{miraghaei2017nuclear} or may be automatically included in FRI samples by definition. Therefore, they are not included in our analysis. The detection of restarting sources shows that the radio emission of the core could be due to a new cycle of AGN activity which can produce a larger scale jet rather than a small scale one. If we assume that restarting sources roughly include 20 percent of FRII radio galaxies, the contribution of radio emission of the core out of newborn jets is even lower than this fraction. This is because a potentially extended newborn jet spends shorter time at the core compared to the time it spends on the way of expansion into a large scale jet or lobe.

\begin{figure}[h!]
	\begin{center}
		\includegraphics[width=\columnwidth]{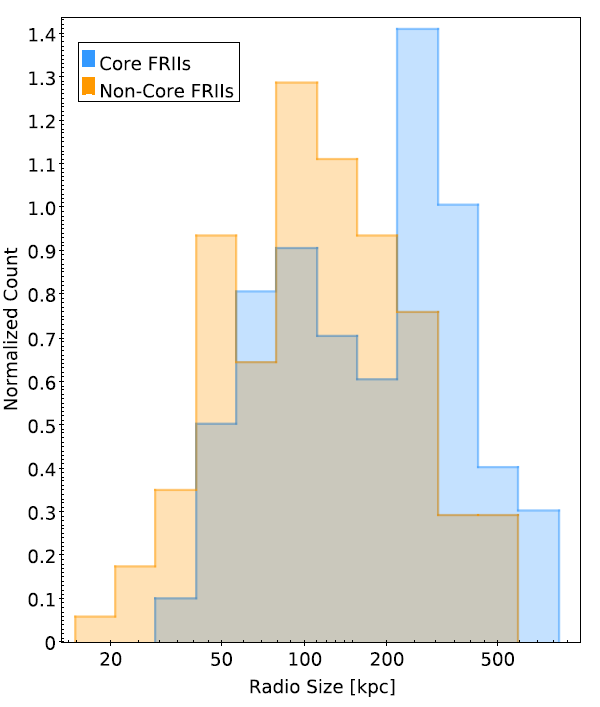}
		\caption{The distributions of radio sizes for the core (blue) and non-core (orange) FRII radio galaxies of LERG type. The histograms represent the non-matched sources from the first subsample and are normalized to unity based on the areas under the plots.}
		\label{fig11}
	\end{center}
\end{figure}

 To explore this more, we compare the distributions of physical sizes of FRIIs in our sample for the core and non-core sources (Fig.\ref{fig11}). The core FRIIs are on average larger compared to the non-core FRIIs consistent with the result presented in \citet{bruni2018probing}. Although we could not distinguish between the radio emissions from small scale jets with those of new born jets, we could slightly modify our results for the observed differences in the physical radio sizes by comparing core and non-core samples with the same distributions of radio sizes. Therefore, both samples show to have the same probabilities for displaying new cycles of AGN activities.

In order to do this, we construct three dimensional matched samples of core/non-core FRIIs of the LERG type in stellar mass-redshift-physical radio size space. The method is the same as described in Section \ref{sec5}. The bin sizes for matched sources are the same as Section \ref{sec5} for the stellar mass and the redshift. Sources with 50kpc differences in radio size ($\Delta$[size]=50kpc) are selected in 3D matched samples. We then compare the properties of matched samples. The 3D results confirm the result in Section \ref{sec5} for the host galaxy properties including [O III] luminosities. We conclude that our results in Table 3 are not biased by the radio size of the sources.

\section{Conclusion and discussion}
We have investigated the core radio emission of AGN using a sample of FRII radio galaxies presented by \citet{miraghaei2017nuclear}. FRIIs with strong, weak or no radio detection at the core (as core or non-core FRIIs) have been compared, based on the optical properties of their host galaxies. Three redshift and radio luminosity cuts have been used to construct complete samples of the core FRIIs and their corresponding non-core FRIIs. Only FRII radio galaxies with radio sizes above 18 arcseconds have been selected to remove the uncertainties which come from the low resolution of the radio maps.
The HERG and LERG FRIIs have been considered separately in order to remove the bias caused by fundamental differences in these two populations of FRII radio galaxies.

Comparison between the two samples shows that core FRIIs of the LERG type have higher [O III] line luminosities than non-core FRIIs of this type. The comparisons have been repeated for two dimensional matched samples of core/non-core FRIIs in the same stellar mass and redshift bins. The significance of these differences has been evaluated by the K-S test which shows differences of the order 2-3$\sigma$ for [O III] line luminosities between the two FRII samples.
No differences between the distributions of core and non-core FRIIs have been detected in the $M_{BH}$, concentration, colour, $D_{4000}$, SFR and galaxy size when the matched samples have been considered. 

The core and non-core samples of HERG FRIIs show no significant differences, probably due to the small sample sizes. Therefore, a larger sample is needed to investigate the differences.

We investigated the [O III] luminosity further and showed that this property is weakly correlated to the total radio luminosity in both FRII LERG and FRII HERG  samples. In terms of core radio luminosity, a weak correlation is detected for the LERG sample while [O III] luminosities in HERGs seem to be anti-correlated to the radio luminosities of the core. These results may be affected by redshift.

We also investigated the core and non-core FRIIs in the [O III] luminosity-$D_{4000}$ plane and showed that core FRIIs have higher [O III] luminosities compared to non-core FRIIs in either $D_{4000}$$>$1.7 or $D_{4000}$$<$1.7 for LERGs. This indicates that the star formation activity in these galaxies does not have a large contribution to the observed difference between the core and non-core FRIIs.
The availability of gas at the vicinity of the central black hole may play a role on the radio power of the core and [O III] luminosities in FRII radio galaxies. A larger gas content provides sufficient fuel into the SMBH to form small scale radio jets as well as higher [O III] luminosities. Thus, the [O III] luminosities are the result of ionization in these regions by the AGN radiation.
Another possibility is ionization of this gas due to the interaction with radio jets. This is investigated and discussed in the literature as jet-driven outflows \citep{o2021compact}.

There is no difference between the two samples for HERGs in the [O III] luminosity-$D_{4000}$ plane. Moreover, the [O III] luminosity and $D_{4000}$ are correlated in HERGs in both samples. Thus, HERGs and LERGs generally show different behaviour when the relation between [O III] luminosity and core radio luminosity is considered. This may originate from the difference in nature, as HERGs are in higher accretion rate regime and display significantly higher [O III] luminosity compared to the LERGs.

Finally, we argued that less than 20 percent of our sample are contaminated by relativistic beaming effect. We modified the bias which comes from the contribution of recurring sources in core and non-core samples by matching the samples in radio source size. The radio-size-matched samples replicated the previous results. We note that our results are based on the sensitivity and resolution of the NVSS, FIRST and VLASS, which limits our ability to see core emission at kpc scale. Since AGN cores may consist of different radio components with different physical sizes, higher resolution radio maps could change some of the results.
This could be better investigated when higher resolution radio maps or VLBI observations become available.

\bibliographystyle{unsrtnat}
\bibliography{ref}  %%% Uncomment this line and comment out the ``thebibliography'' section below to use the external .bib file (using bibtex) .

%%% Uncomment this section and comment out the \bibliography{references} line above to use inline references.
% \begin{thebibliography}{1}

% 	\bibitem{kour2014real}
% 	George Kour and Raid Saabne.
% 	\newblock Real-time segmentation of on-line handwritten arabic script.
% 	\newblock In {\em Frontiers in Handwriting Recognition (ICFHR), 2014 14th
% 			International Conference on}, pages 417--422. IEEE, 2014.

% 	\bibitem{kour2014fast}
% 	George Kour and Raid Saabne.
% 	\newblock Fast classification of handwritten on-line arabic characters.
% 	\newblock In {\em Soft Computing and Pattern Recognition (SoCPaR), 2014 6th
% 			International Conference of}, pages 312--318. IEEE, 2014.

% 	\bibitem{hadash2018estimate}
% 	Guy Hadash, Einat Kermany, Boaz Carmeli, Ofer Lavi, George Kour, and Alon
% 	Jacovi.
% 	\newblock Estimate and replace: A novel approach to integrating deep neural
% 	networks with existing applications.
% 	\newblock {\em arXiv preprint arXiv:1804.09028}, 2018.

% \end{thebibliography}

\end{document}